\newif\ifsingle

\ifsingle
\documentclass[11pt,draftclsnofoot, onecolumn]{IEEEtran}		
\else		
\documentclass[10pt,final, twocolumn]{IEEEtran}
\fi

\usepackage{booktabs} 
\usepackage[table,xcdraw]{xcolor} 
\usepackage{colortbl} 
\usepackage{times}
\usepackage{amsmath,dsfont}
\usepackage{amssymb,amsthm}
\usepackage{epsfig,verbatim}
\usepackage{subcaption}
\usepackage{setspace}
\usepackage{color}
\usepackage{cite}
\usepackage{epstopdf}
\usepackage{graphics}
\usepackage{accents}
\usepackage{acronym}
\usepackage[bookmarks,colorlinks]{hyperref}
\usepackage{booktabs}
\usepackage{mathtools}
\usepackage{float} 
\usepackage{algpseudocode}
\usepackage{enumitem,multirow, multicol}
\usepackage{bm}
\usepackage{colortbl}
\usepackage{pgfplots}
\usepackage{pgfplotstable}
\usepackage{tikz}
\usetikzlibrary{shapes.geometric, positioning, patterns}
\pgfplotsset{compat=1.18}
\usepackage{xcolor}

\usepackage[all=normal,paragraphs=tight,floats=normal,mathspacing=normal,wordspacing=tight,charwidths=tight,mathdisplays=normal,leading=normal]{savetrees}

\usepackage[ruled,linesnumbered,vlined]{algorithm2e}
\SetKwInput{KwData}{\textbf{Init}} 

\DeclareMathAlphabet\mathbfcal{OMS}{cmsy}{b}{n}

\newcommand{\sbrackets}[1]{\left[#1\right]}

\newcommand{\abs}[1]{\left\lvert#1\right\rvert}
\newcommand{\norm}[1]{\left\|#1\right\|}

\newcommand{\expecteds}[1]{\mathds{E}\sbrackets{#1}}

\let\oldnl\nl
\newcommand{\nonl}{\renewcommand{\nl}{\let\nl\oldnl}}

\newcommand{\Capacity}{C}
\newcommand{\Enc}{f_{e}}
\newcommand{\Dec}{f_{d}}
\newcommand{\CB}{\mySet{Q}}

\newcommand{\myVec}[1]{{\boldsymbol{#1}}}
\newcommand{\myMat}[1]{{\boldsymbol{#1}}}

\newcommand{\mySet}[1]{\mathcal{#1}} 

\newcommand{\vx}{\myVec{x}}
\newcommand{\ve}{\myVec{e}}

\newcommand{\vz}{\myVec{z}}

\newcommand{\figWidth}{\columnwidth}
\acrodef{mse}[MSE]{mean-squared error}
\acrodef{mmse}[MMSE]{{minimum mean-squared error}}
\acrodef{mle}[MLE]{maximum likelihood estimation}
\acrodef{snr}[SNR]{signal-to-noise ratio}

\acrodef{ai}[AI]{Artificial Intelligence}
\acrodef{ml}[ML]{{machine learning}}
\acrodef{dl}[DL]{{deep learning}}
\acrodef{nn}[NN]{{neural network}}
\acrodef{dnn}[DNN]{{deep neural network}}
\acrodef{iot}[IoT]{Internet of Things}
\acrodef{rnn}[RNN]{{recurrent neural network}}
\acrodef{cnn}[CNN]{{convolutional neural network}}
\acrodef{dcnn}[DCNN]{{deconvolutional neural network}}
\acrodef{lstm}[LSTM]{{long short-term memory}}
\acrodef{gru}[GRU]{{gated recurrent unit}}
\acrodef{fc}[FC]{{Fully Connected}}
\acrodef{gt}[GT]{{Ground Truth}}
\acrodef{mlp}[MLP]{multi-layer perceptron}
\acrodef{cae}[CAE]{Compressive Auto-encoders}
\acrodef{lbg}[LBG]{Linde–Buzo–Gray}
\acrodef{artoveq}[ARTOVeQ]{Adaptive Rate Task-Oriented Vector Quantization}

\acrodef{sgd}[SGD]{stochastic gradient descent} 

\acrodef{vq}[VQ]{Vector Quantization}
\acrodef{vqvae}[VQ-VAE]{Vector Quantization Variational Autoencoder}
\acrodef{adc}[ADC]{Analog-to-Digital Converter}

\definecolor{Gray}{gray}{0.9}
\definecolor{LightCyan}{rgb}{0.88,1,1}

\IEEEoverridecommandlockouts
\title{Remote Inference over Dynamic Links via Adaptive Rate Deep Task-Oriented Vector Quantization }

\author{
\IEEEauthorblockN{Eyal Fishel, May Malka, Shai Ginzach, and Nir Shlezinger
\thanks{ 
 Parts of this work were presented at the 2023 IEEE International Conference on Acoustics, Speech, and Signal Processing (ICASSP) as the paper~\cite{malka2023learning}. 
 E. Fishel, M. Malka, and N. Shlezinger are with the School of ECE, Ben-Gurion University of the Negev, Beer Sheva, Israel (e-mail: \{eyalfish, maymal\}@post.bgu.ac.il; nirshl@bgu.ac.il). 
 S. Ginzach is with Rafael Advanced Defense Systems (email: shaigi@rafael.co.il).
}}}

\begin{document}

\maketitle
\pagestyle{plain}
\thispagestyle{plain}
%
%
\begin{abstract} 
A broad range of technologies rely on {\em remote inference}, wherein data acquired is conveyed over a communication channel for inference in a remote server. Communication between the participating entities is often carried out over rate-limited channels, necessitating data compression for reducing latency. While deep learning facilitates joint design of the compression mapping along with encoding and inference rules, existing learned compression mechanisms are static, and struggle in adapting   their resolution to changes in channel conditions and to dynamic links.
To address this, we propose  \ac{artoveq}, a learned compression mechanism that is tailored for remote inference over dynamic links. \ac{artoveq} is based on designing nested codebooks along with a learning algorithm employing progressive learning. We show that \ac{artoveq} extends to support low-latency inference that is gradually refined via successive refinement principles, and that it enables the simultaneous usage of multiple resolutions when conveying high-dimensional data.  Numerical results demonstrate that the proposed scheme yields remote deep inference that operates with multiple rates, supports a broad range of bit budgets, and facilitates rapid inference that gradually improves with more bits exchanged, while approaching the performance of single-rate deep quantization methods. 
\end{abstract}
%
%
\acresetall
\section{Introduction}\label{sec:intro}

As data demands and data diversity grow, digital communication systems are increasingly embracing collaborative networks designed for reliable and task-specific communication. This trend is particularly evident in next-generation technologies such as the \acl{iot} and autonomous vehicles, where achieving accurate inference over rate-limited communication channels with low latency is essential~\cite{shlezinger2022collaborative}. Task-based (or goal-oriented) communication has emerged as a necessary and innovative solution for remote inference systems~\cite{zou2022goal}, which tend to operate in two distinct stages. The first stage occurs at the edge or sensing device, where acquired data is conveyed over a rate-limited channel after undergoing compression (source coding) and channel coding~\cite{cover1999elements}. The second stage takes place at the receiver, which extracts the information needed for the task,  e.g., classify an image~\cite{chen2024information}.

Separating the processing involved with communicating data from that associated with inference facilitates the design of remote inference systems, and supports implementation on top of existing communication protocols.  However, separation also often comes at the cost of notable overhead in communication resources, leading to excessive latency, which is often a crucial factor~\cite{shisher2024timely}. This downgrade in performance is a result of the inference task being typically very specific, while the data source is encoded such that it can be entirely recovered, regardless of the task at hand~\cite{lu2023semantics}.  
As such, several studies have attempted to bridge this gap in order to facilitate remote inference over-rate limited links. These include task-based quantization~\cite{shlezinger2018hardware,neuhaus2021task}, semantics-aware coding~\cite{agheli2022semantics, kountouris2021semantics}, and goal-oriented communications~\cite{zhang2022goal,di2023goal}. A common characteristic of these works involves encoding the source based on the inference task rather than prioritizing complete signal reconstruction. This approach supports compact representations, which in turn facilitate lower communication latency compared with the separation based designs~\cite{gunduz2022beyond}. 

Designing task-based compression mechanisms based on statistical models tends to be complicated and is limited to simple tasks that can be represented as linear~\cite{shlezinger2018hardware} and quadratic mappings~\cite{Salamtian19task,bernardo2023design}. Yet, data-driven approaches have been  shown to yield accurate remote inference mechanisms for generic tasks with compact representations. This is achieved by leveraging joint learning of the compression mechanism along with a \ac{dnn}-aided inference rule~\cite{xie2021deep, shao2021learning}.  Such designs employ \ac{dnn}-based encoder-decoder architectures, while constraining the latent features to a fixed bit representation via uniform quantization~\cite{jankowski2020wireless,torfason2018towards,balle2020nonlinear},  scalar quantization~\cite{shlezinger2019deep, shlezinger2022deep, danial2024power}, and  vector quantization~\cite{van2017neural,malka2022decentralized}. Such forms of neural compression, which were shown to achieve highly compressed representation of image~\cite{mishra2022deep}, video~\cite{habibian2019video}, and audio~\cite{zeghidour2021soundstream}  (see detailed survey in~\cite{yang2023introduction}), can be naturally converted into remote inference systems. This is achieved by assigning the encoder and decoder to the sensing and inferring devices, respectively,  while training the overall system for the desired inference metric~\cite{li2018auto}.

The majority of \ac{dnn}-aided compression algorithms operate in a static {single-rate manner}. Namely, the encoder maps the sensed data into a fixed-length bit sequence, which is then processed by the decoder module~\cite{mishra2022deep,agustsson2017soft}. In the context of remote inference, this operation induces two notable challenges when  communicating over time-varying links: 
$(i)$ Once trained, the model's compression rate can not be modified, making it difficult for remote inference systems to adapt to changing channel conditions. Consequently, the system must either adopt a worst-case compression rate, increasing latency, or maintain multiple encoder-decoder model's for different rates, adding complexity. 
$(ii)$ Inference only begins after all the compressed features arrive and are decoded at the inferring device, which has to wait for the entire bit sequence representation to be received before it can provide any form of output. These limitations highlight the need for \ac{dnn}-aided remote inference systems that can operate at different rates and perform inference with minimal latency, ideally starting as soon as the first bits are received.

Several studies have proposed \ac{dnn}-aided compression methods that are not subject to $(i)$ and/or $(ii)$, while focusing on task-invariant compression, i.e., when the decoder recovers the sensed data (typically an image). The first family of multi-rate  methods is that of {\em variable-rate} \ac{dnn}-aided compression, which still require the complete bit sequence to be received for decoding (thus still subject to $(ii)$), but can operate with different bit rates~\cite{cai2019efficient,choi2019variable, yang2020variable, yang2021slimmable,lee2022selective,gupta2022user,duan2024qarv,seo2024raqvae}. The encoder and decoder can be designed to operate with different rates by using multi-scale~\cite{cai2019efficient}, conditional~\cite{choi2019variable}, modulated~\cite{yang2020variable}, and slimmable~\cite{yang2021slimmable} encoder-decoder architectures, or alternatively by masking the latent features~\cite{lee2022selective,gupta2022user} or integrating adaptive normalization~\cite{duan2024qarv}. While all these works utilized uniform scalar quantizers for quantization, \cite{seo2024raqvae} proposed a variable rate compression mechanism that uses vector quantization by training an external Seq2Seq model to generate the codebook on demand. 
The second family of multi-rate \ac{dnn}-aided quantization methods is based on {\em progressive} compression~\cite{toderici2015variable,toderici2017full,johnston2018improved,lee2022dpict,hojjat2023progdtd}. This is typically achieved by using \ac{rnn} based encoders~\cite{toderici2015variable,toderici2017full,johnston2018improved}, which at each step reconstruct the input and encode the residual, such that when each \ac{rnn} output is decoded, an additional residual term is obtained. Alternative approaches to \ac{dnn}-aided progressive compression transform the input into a set of features ordered by importance. These features are fed into uniform scalar quantizers, whose output is used by the decoder to recover the input with growing accuracy~\cite{lee2022dpict,hojjat2023progdtd}.

Despite advancements, current multi-rate \ac{dnn}-aided compression methods have several limitations in the context of remote inference. 
Specifically, while multi-rate methods can be adapted to a task-based setting by replacing the decoder with a \ac{dnn}-aided inference rule, existing methods do not extend to a progressive operation. This is partially due to the fact their focus is mostly on the encoder-decoder architecture, employing simple uniform scalar mappings for quantization (with the exception of \cite{seo2024raqvae}, which supports adjustable vector quantizers at the cost of excessive complexity in runtime, and without enabling progressive operation). 
Existing progressive \ac{dnn}-aided compression methods are highly geared towards a non-task-based setting, where the decoder recovers the input, and progressive operation is obtained by gradual compression of additional features and residual terms that are informative of the input. While one can potentially still employ such architecture in a task-based setting by inferring based on the separately recovered input, such separation-based approaches are known to be inefficient in task-based quantization~\cite{chen2024information}.  

In this work, we tackle the aforementioned  gap  by designing \ac{artoveq}, a multi-rate \ac{dnn}-aided remote inference scheme that naturally supports a progressive operation. \ac{artoveq} is based on a remote inference model that uses a trainable adaptive vector quantization, allowing data compression and inference at multiple rates while using the same underlying architecture. Inspired by nested quantization techniques, we introduce a high-resolution quantization codebook that can be successively decomposed into sub-codebooks of lower resolution~\cite{abdi2019nested, equitz1991successive}. The usage of such nested-style learned codes naturally extends to a progressive operation, where compression is carried out as a form of successive refinement~\cite{cheng2005successive}. This approach supports multi-rate quantization with a single codebook, thereby providing an adaptable and economical solution for varying communication environments.   As our focus is on the learned codebook rather than on the encoder and decoder architecture,  \ac{artoveq} can be combined with existing \ac{dnn}-aided compression mechanisms.

Our main contributions are summarized as follows:
\begin{itemize}  
    %
    \item \textbf{Rate-adaptive learned task-oriented vector quantization:}
We propose a multi-rate task-based vector quantizer that extends the established \ac{vqvae} model  \cite{van2017neural}, which supports remote inference with multi-rate learned vector quantization. Our \ac{artoveq} learns a single codebook that subsumes lower-rate codewords, trained via  a progressive learning scheme~\cite{du2021progressive} that ensures  the remote user (decoder) can reliably infer at various rates, providing a sense of adaptability and reliability.
    \item \textbf{Mixed resolution implementation:} To support a broad range of bit rates with fine granularity we formulate a mixed-resolution implementation of the task-based quantizer. In this model, different features are quantized with different subsets of the learned codebook, thus spanning various different bit rates, while having the encoder learn to map relevant information into features that are quantized with higher resolution. 
    \item \textbf{Progressive quantization via successive refinement:} We extend the learned multivariate codebook of \ac{artoveq} to represent nested codewords, that are naturally applicable in a progressive manner. This allows the multi-rate \ac{dnn}-aided remote inference system to provide predictions of the task with the first codeword received, while gradually improving its reliability with each incoming bit. 
    \item \textbf{Extensive experimentation:} We extensively evaluate our proposed \ac{artoveq} for remote image classification, using the popular edge-oriented MobilenetV2 architecture~\cite{sandler2018mobilenetv2} for the encoder-decoder. Our experiments, which use the CIFAR-100 and Imagewoof datasets, demonstrate that the proposed scheme results in a single model which for all considered rates approaches the performance of{ multiple single-rate \ac{vqvae} models, each  optimized for a specific rate}, while benefiting from mixed-rate implementation and extending to progressive operation with only a minor performance impact. 
\end{itemize}

 The rest of this paper is organized as follows: Section~\ref{sec:intro} reviews the system model and  some preliminaries; our rate-adaptive remote inference scheme is presented in Section~\ref{sec:rate_adaptive_VQ} and evaluated in Section~\ref{sec:results}. Section~\ref{sec:conclusions} concludes the paper.

Throughout this paper, we use
boldface-uppercase for matrices, e.g., $\myVec{X}$, 
and boldface-lowercase for vectors, e.g., $\myVec{x}$. 
We denote the $j$th entry of vector $\myVec{x}$ and the $(i,j)$th entry of matrix $\myMat{X}$ by $[\myVec{x}]_j$ and $[\myMat{X}]_{i,j}$, respectively. We use $\lVert \rVert$, and $ \expecteds{\cdot}$  for the $\ell_2$ norm, and stochastic expectation, respectively.

\begin{figure*}
    \centering
    \includegraphics[width=1.0\textwidth, trim=0 4 0 0cm, clip]{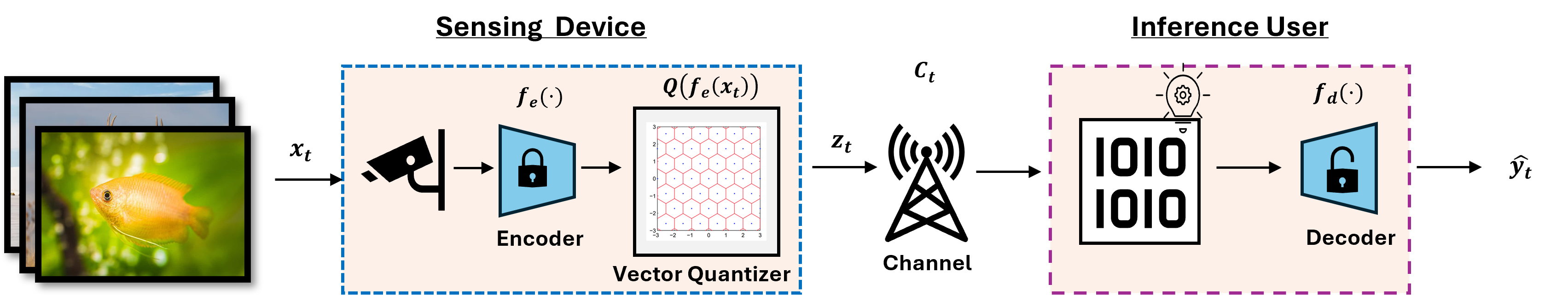}
    \caption{Remote inference system illustration}
    \label{fig:remote_inference_model}
\end{figure*}

%
\vspace{-0.2cm}
\section{System Model and Preliminaries}\label{sec:Model}
\vspace{-0.1cm}
In this section, we review some essential preliminaries and present the system model under consideration. We begin by reviewing basic quantization principles   in Subsection~\ref{ssec:quantization}. Then, we formulate our remote inference problem in Subsection~\ref{ssec:problem_formulation}, and discuss existing mechanisms for \ac{dnn}-aided remote inference  in Subsection~\ref{ssec:remote_inference}.

\subsection{Quantization}\label{ssec:quantization} 


Quantization is concerned with the representation of a continuous-valued signal using a finite number of bits~\cite{gray1998quantization}. The discrete representations produced through quantization should in general  effectively represent signals, even at low resolution, while  maintaining acceptable reconstruction performance. 

Formally, a quantizer, denoted by $\CB_S^{n,k} (\cdot)$ is a mapping from continuous-valued inputs in $\mathbb{R}^n$ into discrete-valued outputs  in $\CB \subset \mathbb{R}^k$ using $\log_2 S$ bits. The set $\CB$, whose cardinality is $|\CB| = S$, represents the {\em quantization codebook}. {This codebook defines the set of possible discrete outputs, forming the basis for the two-stage quantization mapping:} 
Initially, an encoding function  maps the continuous input $\myVec{x} \in \mathbb{R}^n$ into a discrete set $ \{1,2,\ldots, S\}$. Then, a decoding function  maps each item in this discrete set into an associated codeword. Conventionally, $n=k$, and the codeword constitutes a reconstruction of the input. However, in {\em task-based quantization},  the codeword represents some desired information that must be extracted from the input, and thus $k$ can differ from $n$ \cite{shlezinger2018hardware}. When $n=1$, the quantizer is  {\em scalar}, while  $n>1$ denotes a {\em vector quantizer}.


\subsection{Problem Formulation}\label{ssec:problem_formulation}  
We consider a remote inference setting comprised of a sensing device and an inferring user. { At time $t$, the sensing device captures an input data sample $\vx_{t}$, which is conveyed to the inferring user for providing a prediction  $\hat{y}_{t}$}.  For instance, $\vx_t$ can represent an image captured at a remote camera, while $\hat{y}_t$ is the predicted class of the content of the image. The users communicate over a rate-limited channel which is modeled as a bit-pipeline with time-dependent capacity, denoted as $\Capacity_t$, {measuring bits per time unit~\cite{malka2022decentralized}}.  Consequently,  the latency required to transmit $B_t$ bits  at time $t$ is given by $\tau_{t} = \frac{B_t}{\Capacity_{t}}$. The  system is illustrated in Fig.~\ref{fig:remote_inference_model}.  

For conveying  $\vx_{t}$, a quantization mechanism is employed, consisting of: 
$(i)$ an encoder  at the sensing device, denoted $\Enc(\cdot)$,  that maps $\vx_t$ into a $B_t$ bits representation denoted $\vz_t $; 
and $(ii)$ a decoder $\Dec(\cdot)$ implementing a decision rule at the inferring user  that outputs  $\hat{y}_{t}$ based on $\vz_t$. It is assumed that the sensing device knows the current channel capacity $\Capacity_t$, and that the capacity has some lower bound $\Capacity_{\min} > 0$.

We focus on a data-driven setting. During design, one has access to a data set consisting of labeled examples $\mathcal{D} = \{(\vx_i, y_i)\}_{i=1}^{N}$, that is, $N$ pairs of inputs and desired outputs for design purposes. Our goal is to design a remote inference system  based on two performance measures: 
\begin{enumerate}[label={\em P\arabic*}]
    \item \label{itm:accuracy} {\em Accuracy} of the predictions $\hat{y}_t$, where we specifically focus on classification tasks;
    \item \label{itm:latency} {\em Latency} of the inference procedure,  which we constrain to be at most $\tau_{\max}$ (with $ \tau_{\max} \geq \frac{1}{C_{\min}})$.
\end{enumerate}
In principle, the sensing device can be designed to carry out the complete inference procedure. However, we concentrate on the common setting in which only partial pre-processing can be applied due to, e.g., hardware limitations~\cite{shlezinger2022collaborative}.



\subsection{DNN-Aided Remote Inference}
\label{ssec:remote_inference}

A natural approach to design  data-driven remote inference system is to partition a \ac{dnn} suitable for the task at hand  between the sensing and inferring devices, resulting in a trainable encoder-decoder model~\cite{kang2017neurosurgeon}. However, compressing the latent representation using a finite number of bits poses a problem owing to the non-differentiable nature of continuous-to-discrete mappings, and the desire to adjust the bit rate based on channel variations to meet latency constraints. This limits the ability to jointly learn the encoder and decoder mappings using conventional gradient-based deep learning tools. As such, various solutions have been proposed, including modeling scalar quantizers as additive noise during training~\cite{jankowski2020wireless,balle2020nonlinear}, and  soft-to-hard  approximations~\cite{shlezinger2022deep,agustsson2017soft}.   

\begin{figure}
    \centering
    \includegraphics[width=1.0\linewidth]{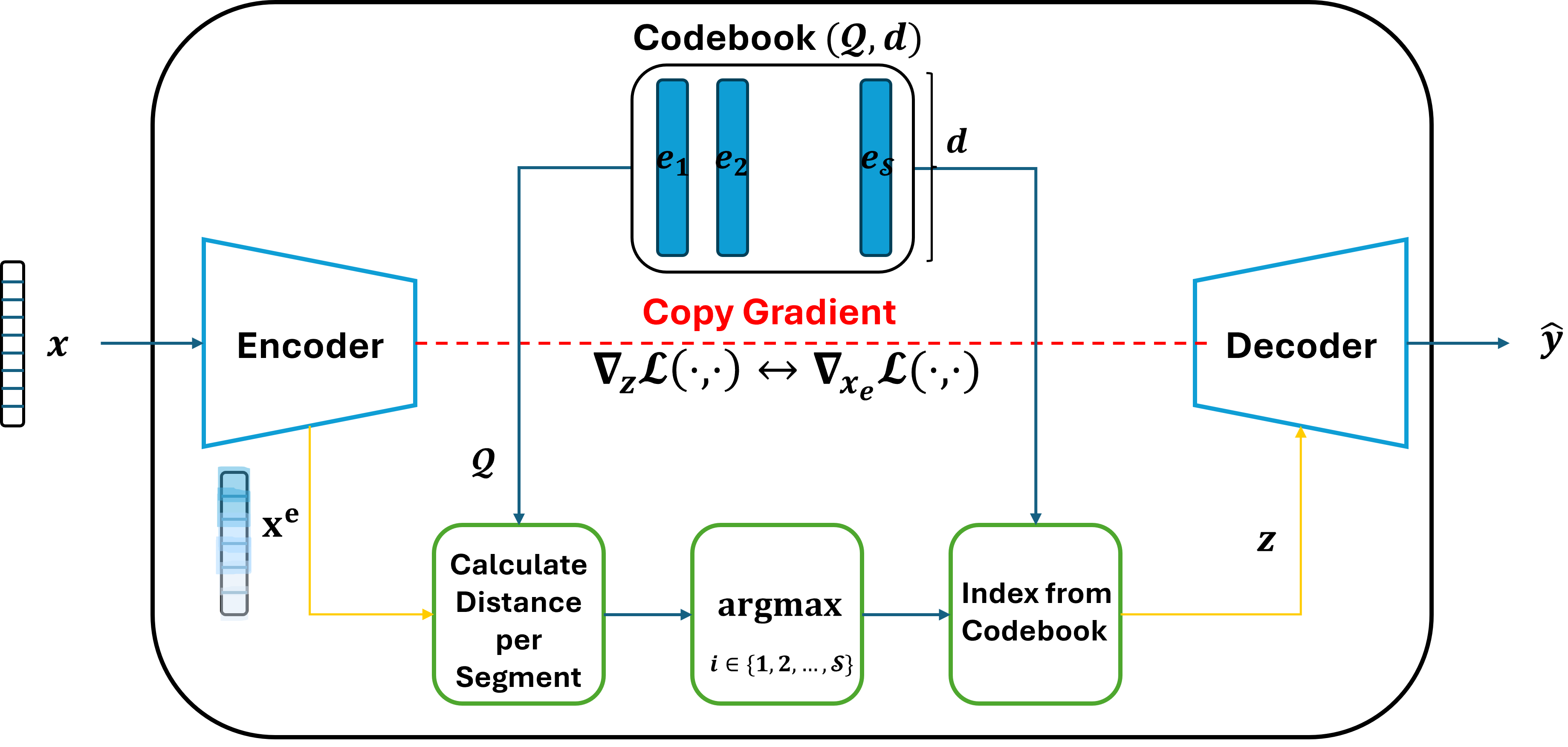}
    \caption{VQ-VAE architecture. The encoder maps the input $\vx$ into the features $\vx^{\rm e}$, which is divided into $M$ sub-vectors of size $d\times 1$. Each sub-vector undergoes  the vector quantization mechanism, which selects an embedding based the distance from the codebook vectors. The decoder is applied to the collection of quantized sub-vectors for inference. 
    }
    \label{fig:VQ-VAE}
\end{figure}

Another approach to bypass the non-differentiable step is to use straight-through gradient estimators. A well-known example of this approach is the well-established \ac{vqvae}~\cite{van2017neural}, illustrated in Fig.~\ref{fig:VQ-VAE}. Gradients are passed through the quantization step, allowing for joint learning of the encoder, codebook, and the decoder, despite the non-differentiable nature of the quantization. This joint optimization forms the foundations for the \ac{vqvae} model, which consists of three components: a \ac{dnn} encoder, $\Enc(\cdot)$, a quantization codebook, $\CB$, and  a \ac{dnn} decoder $\Dec(\cdot)$. The codebook $\CB$ is comprised of $\abs{\CB} = S$ vectors of size $d$. The input sample, $\vx_{t}$, is processed by the encoder into $\vx_{t}^{\rm e} = \Enc(\vx_{t})$ which serves as a low-dimensional representation of the input. Subsequently, the vector $\vx_{t}^{\rm e}$ is decomposed into $M$ vectors of size $d$, denoted $\{\vx_{t,m}^{\rm e}\}_{m=1}^M$, and each is represented by the closest codeword in $\CB$. Thus, the latent representation $\vz_t$ is the stacking of
\begin{equation}
    \label{eqn:quant2ind}
    \vz_{t,m} = \mathop{\arg\min}_{\ve_j\in\CB} \norm{\vx_{t,m}^{\rm e} - \ve_j}_{2}^{2}. 
\end{equation}
  The  quantized $\vz_{t}$ is processed by the decoder into $\hat{y}_{t} = \Dec(\vz_{t})$,  and the number of bits conveyed is $B_t = M \cdot \log_2 S$. 
  
  To jointly train the encoder-decoder while learning the codebook $\CB$, the  \ac{vqvae} uses a loss function comprised of three terms as follows: 
\begin{align}
\mathcal{L}_{\rm tot}({y}_t ; \vx_t) =& \mathcal{L}(y_t ; \hat{y}_t) + \norm{{\rm sg}\left(\vx_t^{\rm e}\right)-\vz_t}_{2}^{2} 
\notag \\
&
+ \beta\norm{\vx_t^{\rm e}-{\rm sg}\left(\vz_t\right)}_{2}^{2},
\label{eqn:train_loss}
\end{align}
where ${\rm sg}(\cdot)$ is the stop-gradient operator.
The first term  in ~\eqref{eqn:train_loss},  $\mathcal{L}\big(y_t;\hat{y}_t\big)$, is the task-dependent loss, (e.g., cross entropy for classification).  The second term is the VQ-loss, which  moves the codebook vectors closer to the encoder outputs. The third term is the commitment loss, which causes the encoder outputs to be similar to the codebook vectors. The hyperparameter  $\beta >0$ balances the influence of the commitment loss on 
$\mathcal{L}_{\rm tot}$. 
While alternative loss measures have been recently proposed for training the \ac{vqvae} to boost improved utilization of its codebook~\cite{gautam2023soft,fifty2024restructuring}, \eqref{eqn:train_loss} is to date the common loss used for training such \ac{dnn}-aided vector quantizers, see~\cite{huang2023towards,dong2023peco}.
The loss in ~\eqref{eqn:train_loss} is stated for a given codebook size $S$, resulting in a model that is fixed to a given bit budget $B_t$.

\section{\ac{artoveq}}
\label{sec:rate_adaptive_VQ} 
In this section we introduce the proposed \ac{artoveq}, designed for remote inference over dynamic channels as formulated in Subsection~\ref{ssec:problem_formulation}. We commence by detailing its high level rationale in Subsection~\ref{sesc:adaptive_VQ_rationale}, after which we present its trainable rate-adaptive vector codebook in Subsection~\ref{ssec:rate_adaptive_mech}. We then show in Subsections~\ref{ssec:identical_resolution}-\ref{ssec:successive_refinement} how the design of \ac{artoveq} naturally extends to support multi-rate and progressive quantization, respectively, with a single codebook. We conclude with a discussion provided in Subsection~\ref{ssec:discussion}   
%
%
\subsection{High Level Rationale}
\label{sesc:adaptive_VQ_rationale} 
The \ac{vqvae} algorithm of ~\cite{van2017neural}, recalled in Subsection~\ref{ssec:remote_inference}, can be used for high performance remote inference (in the sense of \ref{itm:accuracy}) when employed over a static channel (in which the capacity and latency constraints, dictating the bit budget $B_t$, are fixed), owing to its ability to learn task-oriented vector quantization codebooks. Nonetheless, its application for remote inference is not suitable for dynamic channels, as it cannot adapt its bit rate to the the channel conditions. Moreover, its operation is non-progressive, i.e., the decoder needs to receive all bits representing the codeword for inference, which limits its minimal inference latency (\ref{itm:latency}). 

Our proposed \ac{artoveq} builds on the ability of \ac{vqvae} to learn task-oriented multi-resolution codebooks, while overcoming its lack of flexibility and progressiveness by handling a codebook that accommodates multiple-bit resolutions. This is achieved by incorporating the following aspects:
\begin{enumerate}[label={\em A\arabic*}]
    \item \label{itm:UniversalCodebook} A single codebook $\CB$ is designed to support all multi-level bit resolutions by restricting it to be decomposable into sub-codebooks that are used for reduced bit rates. 
    \item \label{itm:Training} A dedicated training algorithm is proposed, which combines principled initialization for vector quantization  based on the  \ac{lbg} algorithm~\cite{linde1980algorithm}, alongside a gradual learning mechanism that allows the same decoder to be reused with all sub-codebooks. 
    \item \label{itm:nestedCodebook} By further restricting the learned codebook to take the form of nested vector quantization~\cite{abdi2019nested}, we enable a progressive operation, where on each incoming bit the decoder can successively refine its predication.
\end{enumerate}
In the following subsections we design \ac{artoveq} by gradually incorporating \ref{itm:UniversalCodebook}-\ref{itm:nestedCodebook} into its design. 


\begin{figure*}[t]
    \centering
    \includegraphics[width=1.0\linewidth]{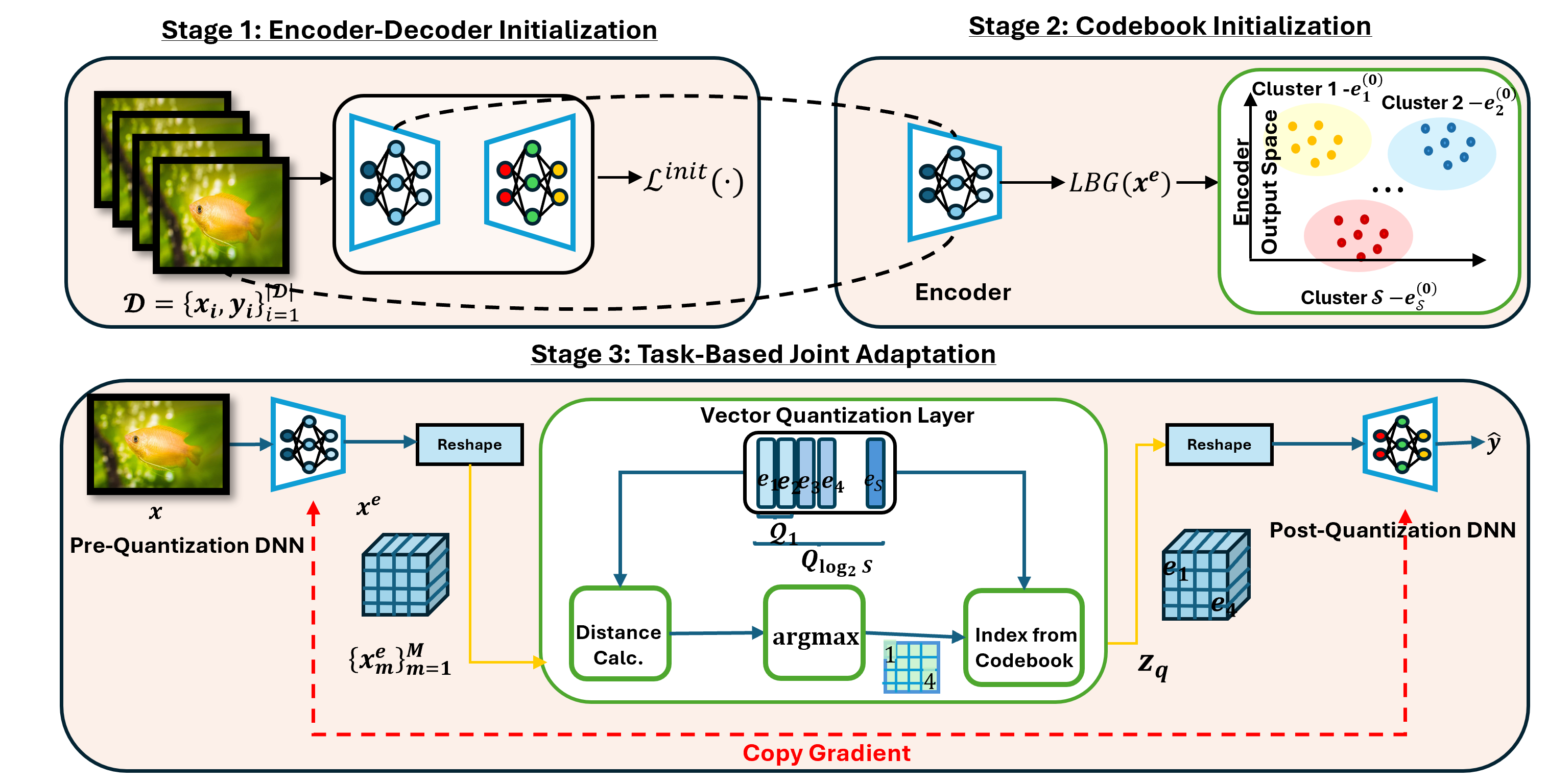}
    \caption{ARTOVeQ training illustration} 
    \label{fig:arch_ART0VeQ}
\end{figure*}

\subsection{ Rate-Adaptive Learned Vector Codebook}
\label{ssec:rate_adaptive_mech} 
Here, we design a \ac{vqvae}-based architecture that enables multi-rate vector quantization, thus meeting \ref{itm:UniversalCodebook}, and present  a training algorithm that enables rate adaptive task-oriented quantization following \ref{itm:Training}.

\subsubsection{Architecture}
Using the  \ac{vqvae} architecture  outlined in Subsection~\ref{ssec:remote_inference}, which is generic in the sense that it is invariant of the specific \acp{dnn} used for the encoder and decoder, we construct a single codebook that accommodates multiple resolutions by iteratively doubling its number of codewords. Specifically, for each quantization level, $l = 1,2,\ldots \log_2 S$, up to some maximum compression rate $S = |\CB|$, a dedicated codebook is maintained $\CB_l$. The process begins with constructing the 1-bit resolution codebook, followed by the 2-bit resolution codebook, and so forth, until the maximum compression rate is reached. This design guarantees that
\begin{equation}\label{eqn:nested_quantization}
  \CB_1 \subset  \CB_2 \subset\cdots \subset  \CB_{\log_2{S}}.
\end{equation}
From \eqref{eqn:nested_quantization} it follows that the first two codewords are derived from $\CB_1$; the first four are derived from $\CB_2$; and so on. The users thus manage a unified codebook encompassing all quantization resolutions, avoiding storing individual codebooks for each resolution.

As new samples $\myVec{x}_t$ become available to the sensing device, the quantization level is initially determined using
\begin{equation} \label{eqn:latency_constraint}
    l_t = \operatorname{max} \bigg \{ l\in \{1,2, \ldots \log_2 S\} \bigg| \dfrac{l}{M \cdot C_t} \leq \tau_{\operatorname{max}} \bigg \}. 
\end{equation} 
After determining the quantization level, remote inference is performed on the discrete outputs at the central server.

\subsubsection{Training} 


Given a dataset $\mathcal{D} = \{(\vx_i, y_i)\}_{i=1}^{N}$, the training algorithm sets the encoder $\Enc(\cdot)$, the codebook $\CB$, and the decoder $\Dec(\cdot)$, through a gradual learning process. This approach is organized in three stages, designed to enable the model to operate at progressively higher resolutions over time while retaining previously acquired knowledge.

\smallskip
{\bf Stage 1: Encoder-Decoder Initialization:}
The first training stage uses $\mathcal{D}$ to obtain a warm start for the encoder-decoder configuration $\Enc(\cdot),\Dec(\cdot)$. This is achieved by training both models as a sequential \ac{dnn} without including quantization, i.e., mapping an input $\vx$ into $\Dec(\Enc(\vx))$. In particular, using the task-dependent loss $\mySet{L}$, the empirical risk that guides the initial setting of $\Enc(\cdot),\Dec(\cdot)$ is given by
\begin{equation}
\label{eqn:LossInit}
    \mySet{L}_{\mySet{D}}^{{\rm init}}(\Enc, \Dec) = \frac{1}{|\mySet{D}|}\sum_{i=1}^{|\mySet{D}|} \mySet{L}(\Dec(\Enc(\vx_i)), y_i).
\end{equation}

\smallskip
{\bf Stage 2: Codebook Initialization:}
Next, we initialize the vector codebook $\CB$ with $S$ codewords of size $d\times 1$. To that aim, we first pass every $\vx_i\in \mySet{D}$ in the trained encoder, and divide its output $\vx_i^{\rm e}$ into $M$ sub-vectors of size $d\times 1$, denoted $\{\vx_{i,m}^{\rm e}\}_{m=1}^M$. These sub-vectors are aggregated into a new unlabeled dataset $\mySet{D}_{\rm Q} = \big\{\{\vx_{i,m}^{\rm e}\}_{m=1}^M\big\}_{i=1}^N$.
{The obtained $\mySet{D}_{\rm Q}$ is used to initialize the codebook with $S$ codewords $ \{\ve^{\rm LBG}_k\}_{k=1}^S$ using the \ac{lbg} algorithm~\cite{linde1980algorithm}. The \ac{lbg} algorithm iteratively constructs a codebook for a single rate by creating non-task-based vector quantizers, with the goal of minimizing distortion (measured via the $\ell_2$ norm) in its codeword representation over $\mySet{D}_{\rm Q}$. Specifically, it seeks to minimize the following loss function}:
\begin{equation}
\label{eqn:LossLBG}
    \mySet{L}_{\mySet{D}_{\rm Q}}^{{\rm LBG}}\left(\{\ve_k\}\right) = \frac{1}{|\mySet{D}|}\sum_{\vx^{\rm e} \in\mySet{D}_{\rm Q} } \min_{k=1,\ldots,S} \left\|\vx^{\rm e} - \ve_k \right\|_2.
\end{equation}
This principled codebook initialization facilitates tackling a core challenge in training \acp{vqvae}, i.e., the frequent learning of under-used codewords~\cite{fifty2024restructuring}, without having the alter the \ac{vqvae} loss such that it can be utilized for boosting support of multiple rates in the subsequent stage.


\smallskip
{\bf Stage 3: Task-Based Joint Adaptation:}
The codebook vectors are then jointly updated as learnable parameters, along with the encoder and decoder. In this stage, we sequentially refine the model for each quantization level \( l = 1, 2,  \ldots, \log_2 S \). For each level \( l \), the codebook \( \CB_l \) is constructed by expanding the previous codebook \( \CB_{l-1} \) with additional code vectors, initially drawn from the LBG initialized $\{\ve^{\rm LBG}_k\}_{k=1}^S$.  

Specifically,  when training at step $l \in \{1,\ldots, \log_2 S\}$, one already has a codebook $\CB_{l-1}$ along with the encoder-decoder trained so far. Thus, the extended codebook $\CB_l$ is initialized by setting its first $2^{l-1}$ codewords, denoted $\{\ve_1^{(l)},\ldots, \ve_{2^{l-1}}^{(l)}\}$, to be the same ordering of codewords in  $\CB_{l-1}$, denoted $\{\ve_1^{(l-1)},\ldots, \ve_{2^{l-1}}^{(l-1)}\}$, while setting the remaining  $2^{l-1}$ codewords to be the corresponding  codewords from $\{\ve^{\rm LBG}_k\}_{k=1}^S$.  Then, to learn $\CB_l$ while having the decoder be suitable for all sub-codebooks in $\CB_l$, we further train $\Enc(\cdot)$, $\CB_l$, and $\Dec(\cdot)$ using a loss measure which accounts the inference accuracy achieved with {\em all codebooks} of quantization levels up to $l$, while encouraging the first $2^{l-1}$ codewords $\CB_l$ not to deviate much from those already learned. This loss  at step $l$ is 
\begin{align}
&\mathcal{L}_{\rm tot}^{(l)}({y}_t ; \vx_t) = \sum_{j=1}^{l}\mathcal{L}\left(y_t ; \hat{y}_t^{(j)}\right) + \norm{{\rm sg}\left(\vx_t^{\rm e}\right)-\vz_t^{(j)}}_{2}^{2} 
\notag \\
&\quad
+ \beta_j\norm{\vx_t^{\rm e}-{\rm sg}\left(\vz_t^{(j)}\right)}_{2}^{2} + \eta_l \sum_{k=1}^{2^{l-1}} \norm{\ve_k^{(l)} - \ve_k^{(l-1)}}_{2}^{2}.
\label{eqn:VQVAE loss}
\end{align}
In~\eqref{eqn:VQVAE loss}, $\vz_t^{(j)}$ is the vector obtained by quantizing $\vx_t^{\rm e} = \Enc(\vx_t)$ using the first $2^{j}$ codewords in $\CB_l$, while $\hat{y}_t^{(j)} = \Dec(\vz_t^{(j)})$.

Equation~\eqref{eqn:VQVAE loss} encapsulates the cumulative impact of quantization levels up to $l$.  The first three terms are based on the \ac{vqvae} training loss as in \eqref{eqn:train_loss}, aggregated over all resolutions.
The last term promotes rate adaptability, with the hyperparameter $\eta \geq 0$ governing its impact. 
The overall loss using dataset $\mySet{D}$ is 
\begin{equation}
    \label{eqn:LossOverall}
        \mySet{L}_{\mySet{D}}^{{\rm tot}}(\Enc, \CB, \Dec) = \frac{1}{|\mySet{D}|}\sum_{i=1}^{|\mySet{D}|}\sum_{l=1}^{\log_2 S} \mathcal{L}_{\rm tot}^{(l)}({y}_i ; \vx_i).
\end{equation}
%
A concise depiction of the training algorithm, where mini-batch \acl{sgd} is employed for training in Stages 1 and 3, is presented in Algorithm \ref{alg:VQ_training_LBG}, and the overall procedure is illustrated as Fig.~\ref{fig:arch_ART0VeQ}.

\begin{algorithm}
\caption{\ac{artoveq} Training}
\label{alg:VQ_training_LBG} 
\SetKwInOut{Input}{Input}  
\Input{Dataset $\mathcal{D}$; Bits limit $S$; \\ Loss hyperparameters $\{ \eta_{l} \}$ and $\{ \beta_{j}\}$}

\nonl\textbf{Stage 1: Encoder-Decoder Initialization}\;
\For{${\rm epoch}=0,1,\ldots$}
{
  Randomly divide  $\mathcal{D}$ into $P$ batches $\{\mathcal{D}_p\}_{p=1}^P$\;
      \For{$p = 1, \ldots, P$}{
        Compute loss on $\mathcal{D}_p$ using \eqref{eqn:LossInit}\;
        Update $\Enc$ and $\Dec$ using loss gradient\;
      }
}

\nonl\textbf{Stage 2: Codebook Initialization}\;

Obtain $\mySet{D}_{\rm Q}$ by applying $\Enc(\cdot)$ to each $\vx \in \mySet{D}$\;
Set  $\{\ve^{\rm LBG}_k\}_{k=1}^S$ from $\mySet{D}_{\rm Q}$ using \ac{lbg}\;

\nonl\textbf{Stage 3: Task-based Joint Adaptation}\;
Set $\CB_0 = \emptyset$\;
\For{\text{quantization level} $l \in \{1,2,\ldots,\log_2 S\}$}{
Init $\CB_l$ by adding $\{\ve^{\rm LBG}_k\}_{k=2^{l-1}+1}^{2^{l}}$
 to $\CB_{l-1}$\;
\For{${\rm epoch}=0,1,\ldots$}
{
  Randomly divide  $\mathcal{D}$ into $P$ batches $\{\mathcal{D}_p\}_{p=1}^P$\;
      \For{$p = 1, \ldots, P$}{
        Compute loss on $\mathcal{D}_p$ using  \eqref{eqn:LossOverall}\;
        Update $\Enc$, $\CB_l$, and $\Dec$ using loss gradient\;
      }
}
}
\KwOut {Trained $\Enc$ and $\Dec$; codebook $\CB = \CB_{\log_2 S}$.}
\end{algorithm}

\subsection{Mixed Resolution \ac{artoveq}}
\label{ssec:identical_resolution} 


\begin{figure}
    \centering
    \includegraphics[width=1.0\linewidth]{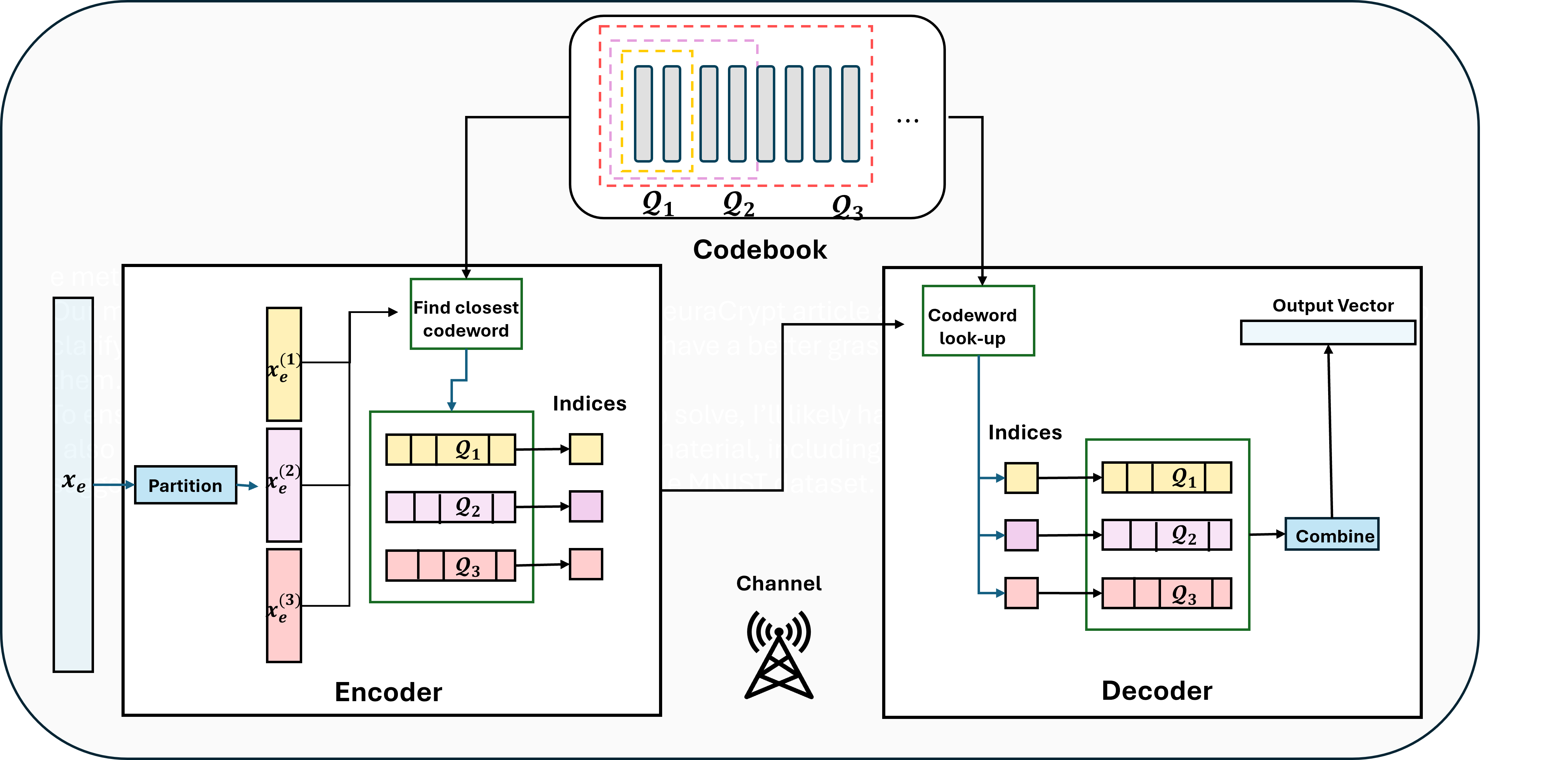}
    \caption{Mixed resolution \ac{artoveq} illustration. Different colors represent different quantization resolutions.}
    \label{fig:mixed_res_ART0VeQ}
\end{figure}
\ac{artoveq} learns a task-oriented vector quantizer using a single codebook that can be applied across multiple resolutions.
Still, once a bit budget $l \in \{1,\ldots,\log_2 S\}$ is fixed, the same $l$-bit codebook is applied to each features sub-vector, at an overall budget at time t of $B_t = M \cdot l$ bits per input. 
However, the fact that the same codebook $\CB$ can be decomposed into multiple codebooks of different resolutions can be leveraged to quantize high dimensional inputs with mixed resolutions applied to different features. 

\subsubsection{Architecture}
To formulate the mixed resolution \ac{artoveq}, 
we recall that  the encoder output $\myVec{x}_t^{\rm e}$ is divided into the $M$ sub-vectors $\{\vx_{t.m}^{\rm e}\}_{m=1}^M$. Each features segment is assigned a specific sub-codebook based on a designated bit resolution $S_m$, with a total bit budget at time $t$ is $B_t = \sum_m \log_2 S_m$ representing the sum of bits allocated across all segments. The resulting bit budget $B_t$ can thus take any value in the range $[M, M+1, \ldots, M \log_2 S]$, indicating that the mixed resolution design provides high bit budget flexibility and granularity, a property not achieved with alternative variable rate learned quantizer architectures whose focus is on the encoder-decoder architecture, e.g., \cite{toderici2015variable,toderici2017full,johnston2018improved,lee2022dpict,hojjat2023progdtd}.  An illustration of the mixed resolution \ac{artoveq} can be seen in Fig.~\ref{fig:mixed_res_ART0VeQ}.  

\subsubsection{Training}
The training of mixed resolution \ac{artoveq} follows the same procedure as in Algorithm~\ref{alg:VQ_training_LBG}, with a slight modification applied in {\bf Stage 3}. Here, instead of progressively increasing the resolution of the codebook and having it employed for quantizing all $M$ features, we gradually increase the resolution of the first $d\times 1$ features $\vx^{\rm e}_{t,1}$, after which we increase the resolution of quantizing $\vx^{\rm e}_{t,2}$, and so on. The rationale in this form of gradual learning draws inspiration from classical image compression methods based on quantizing different components with different resolution, e.g.,~\cite{shusterman1994image}. In doing so, we aim  to consistently have some features quantized with improved resolution, such that the task-based encoder-decoder be encouraged to embed there features that are more informative with respect to the task.

\subsection{Progressive \ac{artoveq}}
\label{ssec:successive_refinement}
%
While the training procedure used by \ac{artoveq} is based on progressive learning, where the resolution of intermediate features gradually grows during training~\cite{du2021progressive}, the resulting quantizer does not immediately support progressive quantization. Specifically, for a chosen bit budget $B_t$, the codewords do not support progressive decoding, namely, the decoder has to have access to all bits representing the compressed features in order to infer. 
Nonetheless, while the formulation of the codebook  in Subsection~\ref{ssec:rate_adaptive_mech} only allows variable-rate operation, the fact that what one learns is the multi-resolution codebook implies that it can naturally extend to have a progressive codebooks, whose codewords incrementally build   on prior representations, as a form of successive refinement. 

\subsubsection{Architecture}
To support progressive quantization, we alter the codebook constraint of \eqref{eqn:nested_quantization} to be one which supports successive refinement of initial low-resolution representations of the codewords. 
Drawing inspiration from nested quantization, which is typically considered in the context of uniform~\cite{abdi2019nested} and lattice codebooks~\cite{zamir2002nested}, we constrain each intermediate codebook $\CB_l$ to represent a one bit refinement of $\CB_{l-1}$. Mathematically, for each $l \in \{1,\ldots, \log_2 S\}$ there exist $d\times 1$ vectors $\tilde{\ve}^{(l)}_1, \tilde{\ve}^{(l)}_2$ such that
\begin{equation}
\label{eqn:ProgConst}
    \CB_{l} = \CB_{l-1}+\left\{\tilde{\ve}^{(l)}_1, \tilde{\ve}^{(l)}_2\right\},
\end{equation}
with $+$  being the Minkowski set sum, thus $|\CB_l| = 2\cdot |\CB_{l-1}|$. 

The constrained codebook form in ~\eqref{eqn:ProgConst} enables progressive recovery via successive refinement. Specifically for an encoder output $\vx^{\rm e}_t$ and its decomposition into $\{\vx^{\rm e}_{t,m}\}$, the decoder only needs one bit per each sub-vector to recover their representation in $\CB_1$ and use it for inference. With the next $M$ bits, the decoder obtains the improved representation in $\CB_2$, and uses it to improve its inference output, and so on.






\subsubsection{Training}
Progressive \ac{artoveq} is based on the learned task-based  vector quantizer detailed in Subsection~\ref{ssec:rate_adaptive_mech}, while introducing an alternative constraint on the learned multivariate codebook in the form of \eqref{eqn:ProgConst}. Accordingly, the training procedure of progressive \ac{artoveq} is based on the learning procedure stated in Algorithm~\ref{alg:VQ_training_LBG}, with three main differences introduced to support the constrained progressive form \eqref{eqn:ProgConst}:
\begin{itemize}
    \item Since the \ac{lbg} algorithm is based on clustering the inputs without accommodating the desired constrained form, the initialization of the codebook in {\bf Stage 2} is omitted.
    \item For each quantization level $l$, the aspects of the codebook that are learned are the two difference vectors $\tilde{\ve}^{(l)}_1, \tilde{\ve}^{(l)}_2$. These are randomly initialized for each resolution.
    \item As the codebooks no longer satisfy $\CB_{l-1} \subset \CB_{l}$, but instead hold \eqref{eqn:ProgConst}, the regularizer encouraging the former in $\mathcal{L}_{\rm tot}^{(l)}$ is canceled, i.e., we set $\eta_l =0$ for each $l$ in \eqref{eqn:VQVAE loss}.
\end{itemize}
The resulting training algorithm is summarized as Algorithm~\ref{alg:successive_refinement_training}.

\begin{algorithm}
\caption{Progressive \ac{artoveq} Training}
\label{alg:successive_refinement_training} 
\SetKwInOut{Input}{Input}  
\Input{Dataset $\mathcal{D}$; Bits limit $S$; \\ Loss hyperparameters $\{ \beta_{j}\}$}

Initialize $\Enc$ and $\Dec$ via {\bf Stage 1}\;

Set $\CB_0 = \emptyset$ and $\eta_l \equiv 0$\;
\For{\text{quantization level} $l \in \{1,2,\ldots,\log_2 S\}$}{
Randomize $\tilde{\ve}^{(l)}_1, \tilde{\ve}^{(l)}_2$\;
\For{${\rm epoch}=0,1,\ldots$}
{
  Randomly divide  $\mathcal{D}$ into $P$ batches $\{\mathcal{D}_p\}_{p=1}^P$\;
      \For{$p = 1, \ldots, P$}{
      Set $\CB_l$ via \eqref{eqn:ProgConst}\;
        Compute loss on $\mathcal{D}_p$ using  \eqref{eqn:LossOverall}\;
        Update $\Enc$, $\tilde{\ve}^{(l)}_1, \tilde{\ve}^{(l)}_2$,  $\Dec$ using loss gradient\;
      }
}
}
\KwOut {Trained $\Enc$, $\Dec$; differences $\{\tilde{\ve}^{(l)}_1, \tilde{\ve}^{(l)}_2\}_{l=1}^{\log_2 S}$.}
\end{algorithm}

\begin{figure*}[t]
  \centering

  \begin{subfigure}[b]{0.20\textwidth}
    \includegraphics[width=\textwidth]{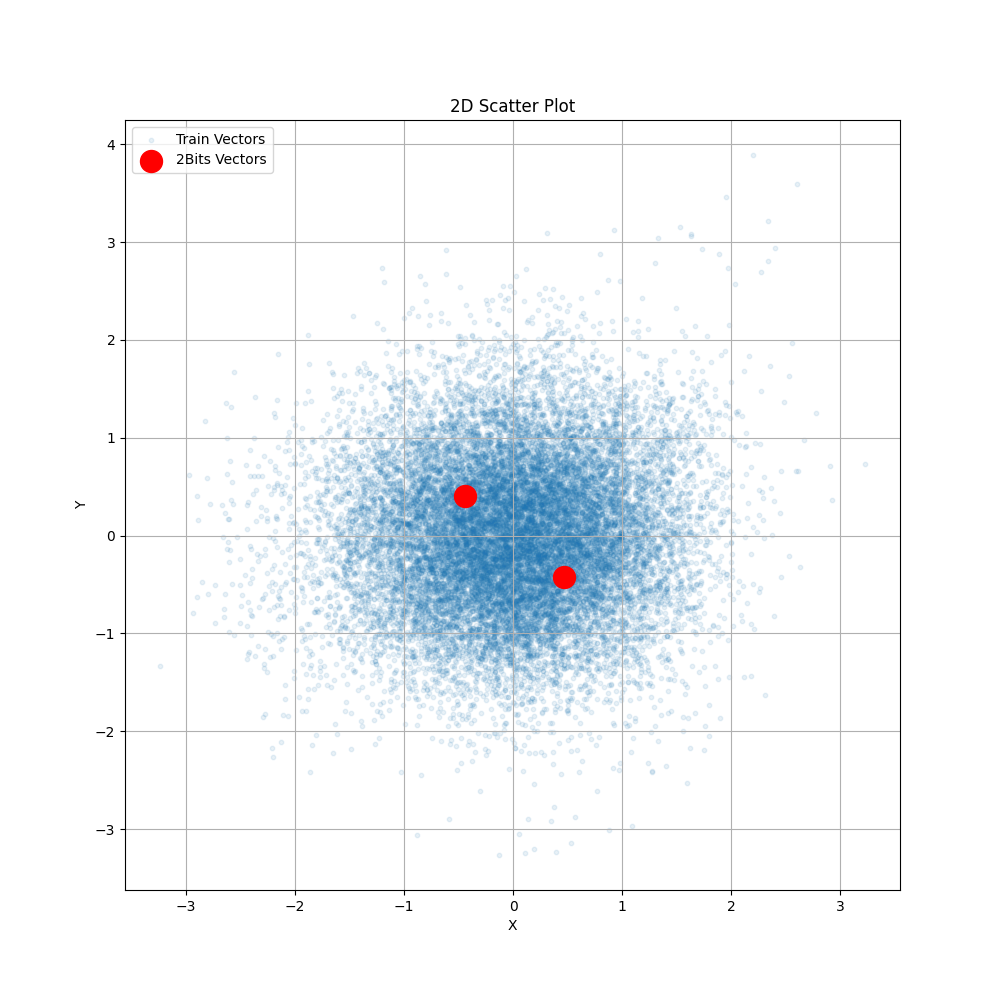}
    \caption{$\CB_1$}
    \label{fig:sub1}
  \end{subfigure}
  \hfill
  \begin{subfigure}[b]{0.20\textwidth}
    \includegraphics[width=\textwidth]{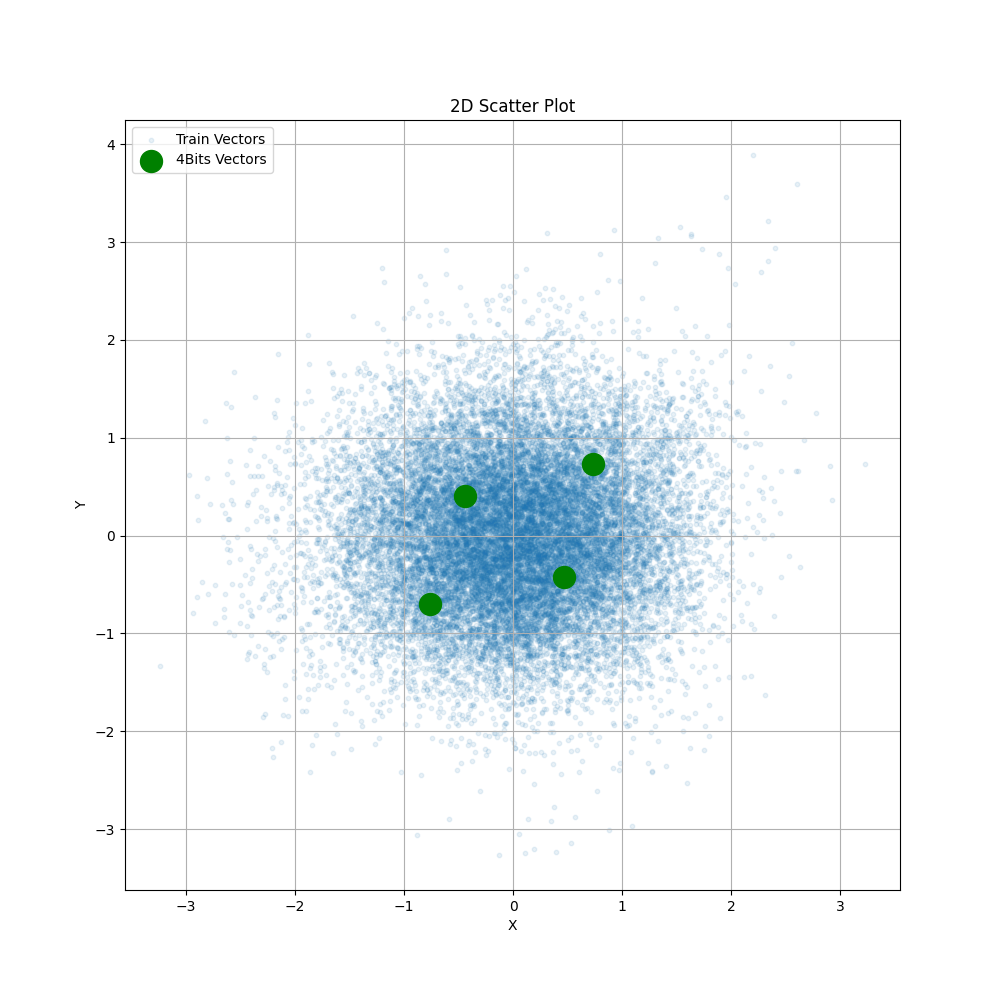}
    \caption{$\CB_2$}
    \label{fig:sub2}
  \end{subfigure}
  \hfill
  \begin{subfigure}[b]{0.20\textwidth}
    \includegraphics[width=\textwidth]{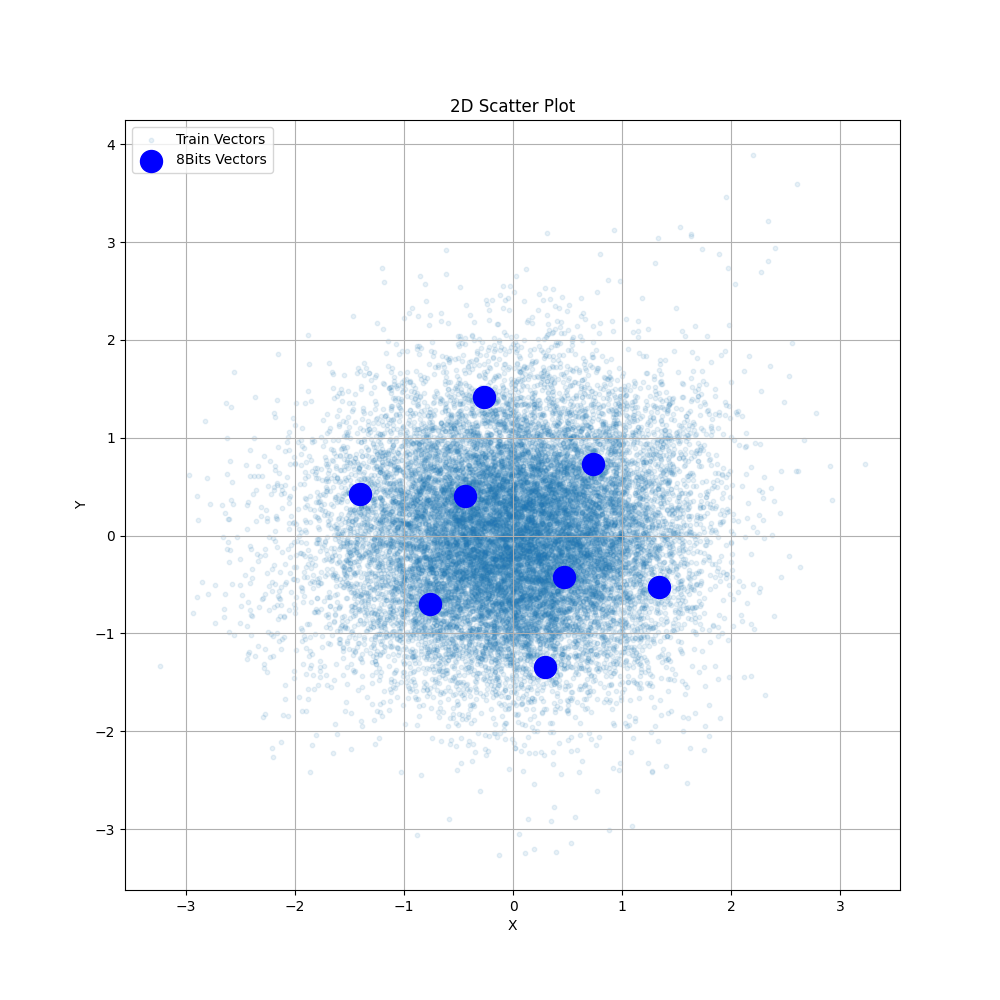}
    \caption{$\CB_3$}
    \label{fig:sub3}
  \end{subfigure}
  \hfill
  \begin{subfigure}[b]{0.20\textwidth}
    \includegraphics[width=\textwidth]{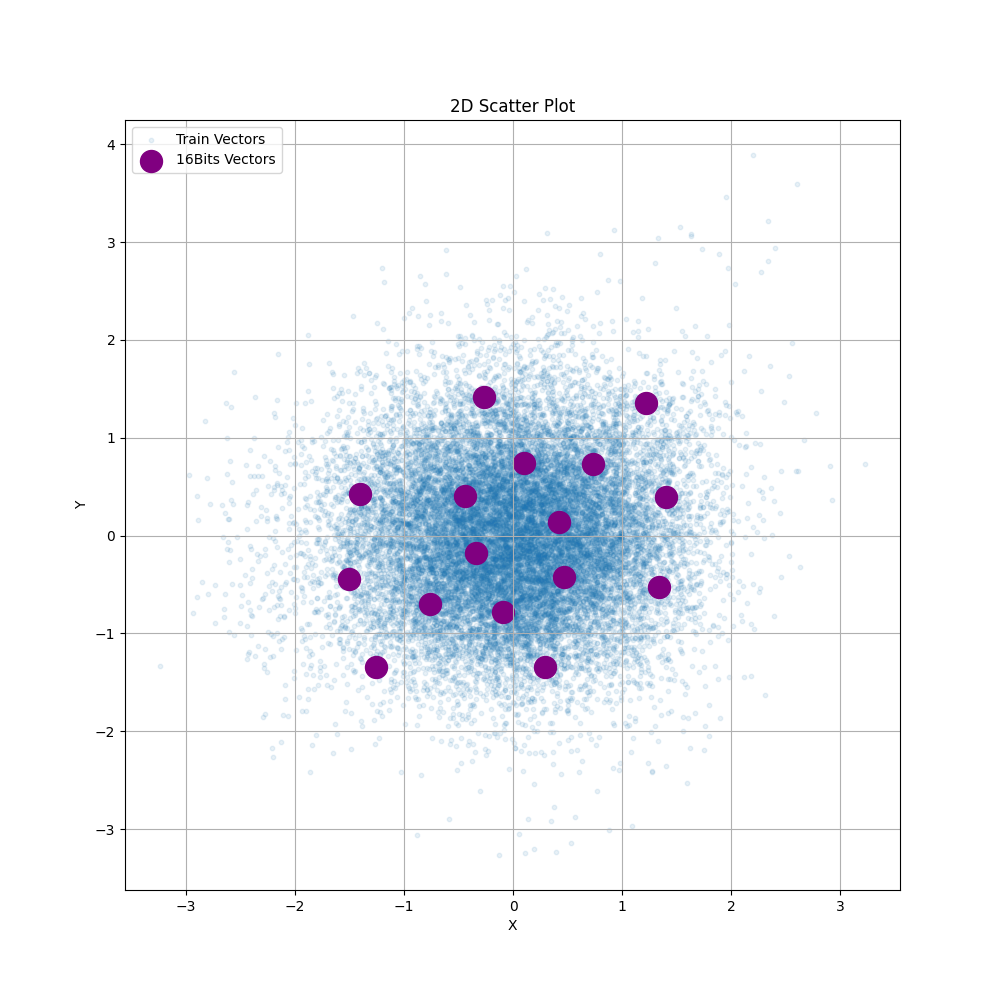}
    \caption{$\CB_4$}
    \label{fig:sub4}
  \end{subfigure}

  \vspace{\baselineskip} 

  \begin{subfigure}[b]{0.20\textwidth}
    \includegraphics[width=\textwidth]{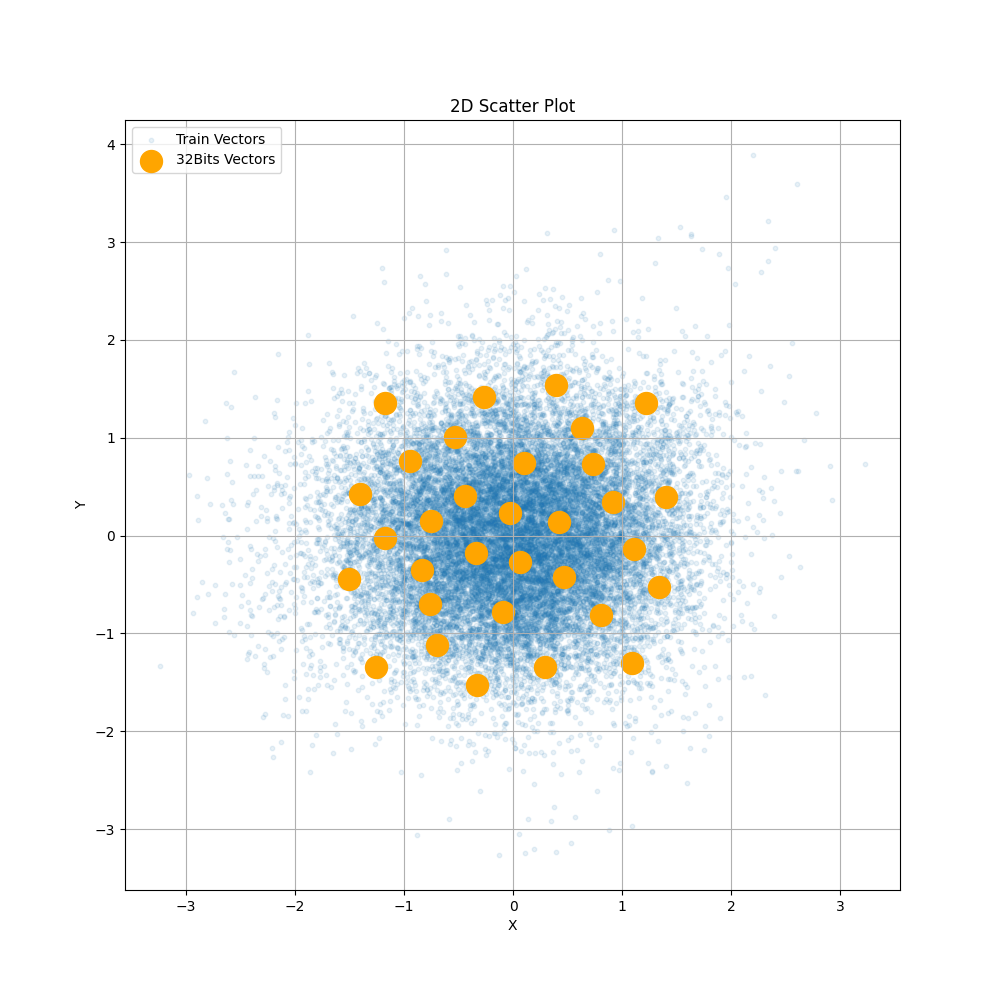}
    \caption{$\CB_5$}
    \label{fig:sub5}
  \end{subfigure}
  \hfill
  \begin{subfigure}[b]{0.20\textwidth}
    \includegraphics[width=\textwidth]{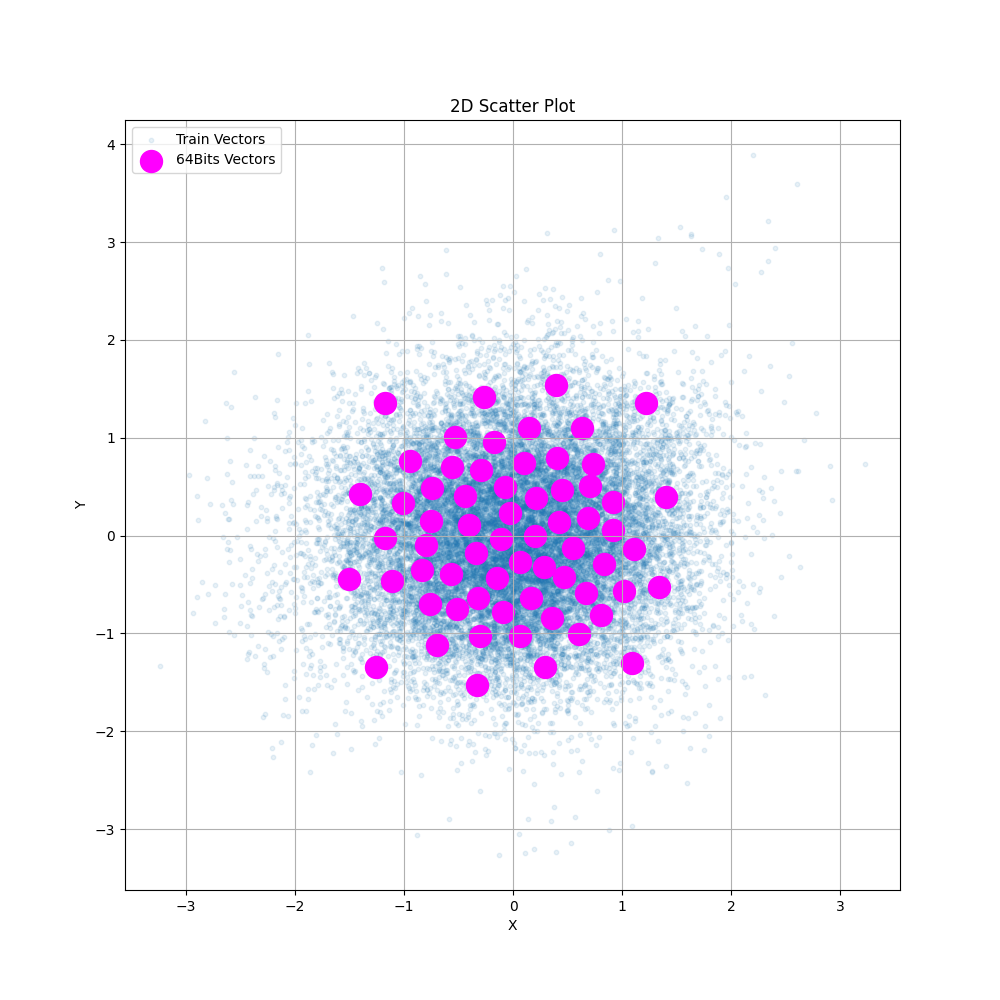}
    \caption{$\CB_6$}
    \label{fig:sub6}
  \end{subfigure}
  \hfill
  \begin{subfigure}[b]{0.20\textwidth}
    \includegraphics[width=\textwidth]{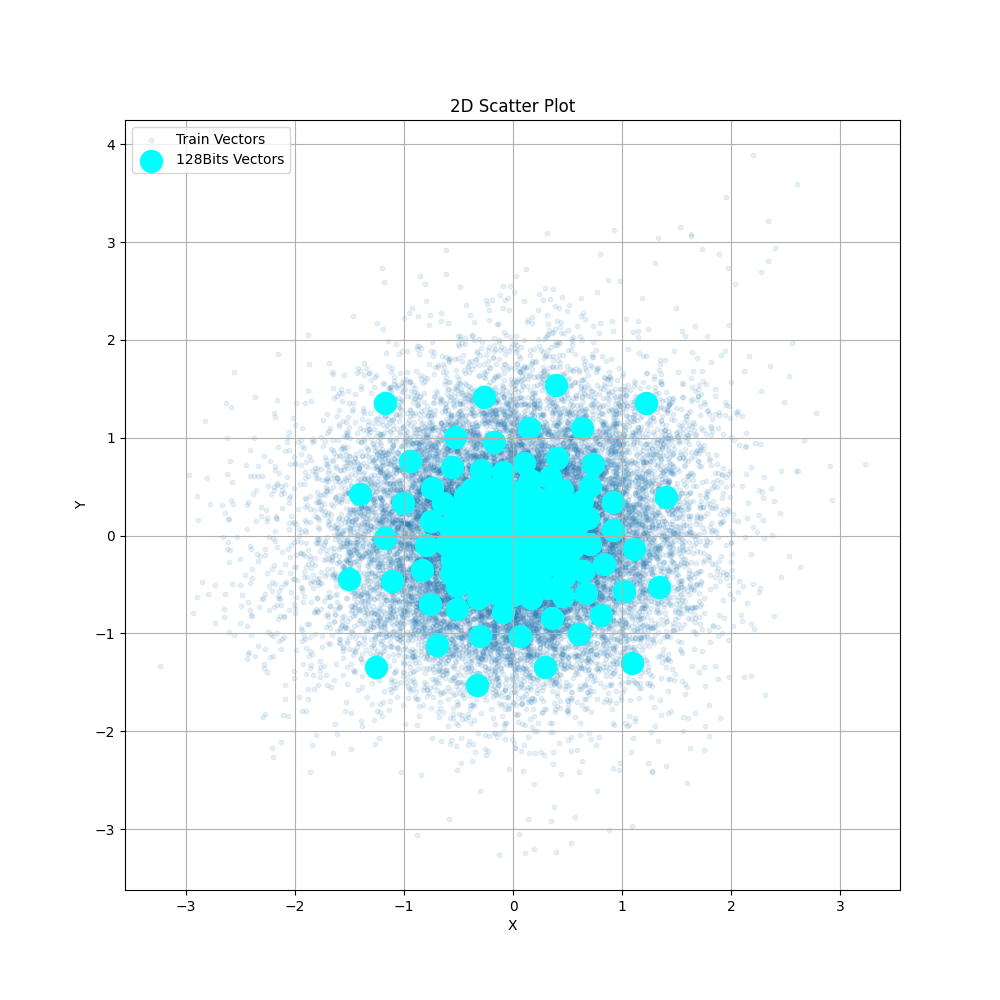}
    \caption{$\CB_7$}
    \label{fig:sub7}
  \end{subfigure}
  \hfill
  \begin{subfigure}[b]{0.20\textwidth}
    \includegraphics[width=\textwidth]{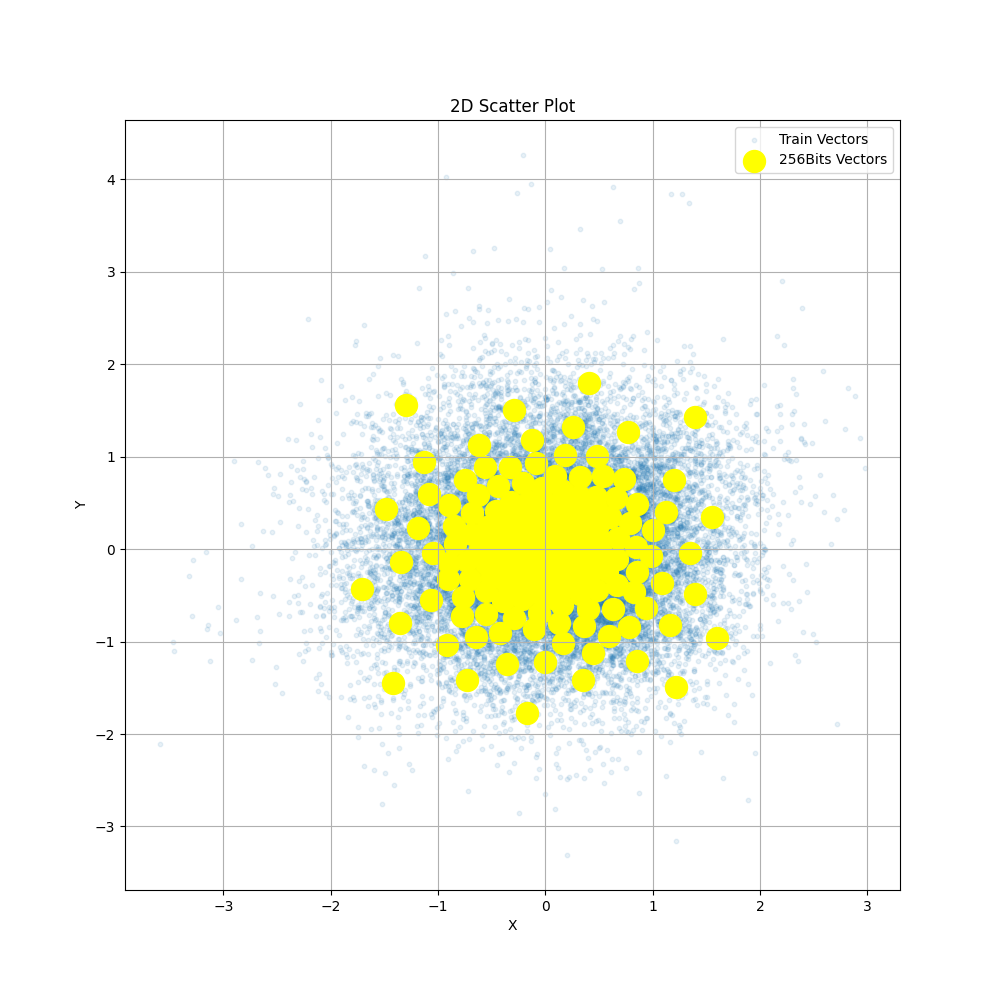}
    \caption{$\CB_8$}
    \label{fig:sub8}
  \end{subfigure}

  \caption{Learned codebook vectors. Embedding dimensions $d=2$}
  \label{fig:learned_cb_vectors}
\end{figure*}

\subsection{Discussion}
\label{ssec:discussion}

The proposed \ac{artoveq} is designed to facilitate learning a single task-based vector quantization codebook that supports finer granularity and progressive decoding through the use of multiple resolution. Accordingly, it is particularly suitable for remote inference over time-varying communication links, e.g., with mobile users~\cite{shlezinger2022collaborative}. This flexibility allows the compression rate to be adjusted according to dynamic channel conditions, ensuring accurate inference, while supporting a broad range of multiple resolutions (via mixed-resolution among different feature sub-vectors), as well as allowing the decoder to provide rapid inference and gradually improve it via successive refinement. Adaptivity is achieved through nested codebooks, and progressive learning techniques, allowing the system to refine its performance over successive iterations or stages of operation. As \ac{artoveq} focuses on learning the quantization codebook and does not restrict the task-based mappings $\Enc$ and $\Dec$, it can be integrated in various \ac{dnn} architectures.


While our setup primarily focuses on a pair of sensing and inferring users, this methodology is extensible to collaborative inference among multiple edge devices. Our approach assumes that the instantaneous channel capacity ($C_t$) is known, enabling the sensing user to determine the appropriate quantization level. However, a potential extension of our scheme could allow it to function without prior knowledge of channel capacity, dynamically tuning the quantization rate during the remote inference process.
Another potential aspect for future exploration,  which stems from the ability to learn a variable rate and progressive vector quantization codebook integrated into a remote inference system, is its ability to enhance data privacy and security. Recent studies have shown that well-designed compression strategies can enhance privacy, providing an additional benefit to our rate-adaptive scheme \cite{alvar2021scalable, lang2023joint}. Furthermore, recent advancements in randomized neural networks demonstrate their potential for ensuring privacy \cite{esfahanizadeh2023infoshape}. While \ac{artoveq} has the potential of supporting such extensions, they would necessitate reformulation of the learning procedure, and are thus left for future investigation.


 
\section{Numerical Experiments}
\label{sec:results} 
In this section, we present the results of our numerical experiments\footnote{The source code and hyperparameters used in this experimental study are available at \url{https://github.com/eyalfish/ARTOVeQ}.}. We first detail our experimental setup in Subsection~\ref{ssec:expSetup}, after which we detail our four main studies, each focusing on a distinct aspect of our approach for image classification: variable-rate task-based compression (Subsection~\ref{ssec:variable_rate}), mixed resolution compression (Subsection~\ref{ssec:mixed_resolution}), progressive compression (Subsection~\ref{ssec:progessive_comp}), and remote inference over dynamic channels (Subsection~\ref{subsec:latency}).

\subsection{Experiential Setup}
\label{ssec:expSetup}
To evaluate our quantization scheme, we use two datasets: CIFAR-100 and Imagewoof. CIFAR-100 consists of 60,000 diverse images with dimensions $3 \times 32 \times 32$, spanning 100 classes and thus encompassing a wide variety of source distributions. Imagewoof contains 10,000 images at a higher resolution of $3 \times 64 \times 64$, but with a slightly narrower set of 10 classes, each representing different dog breeds. This combination allows us to assess our method’s robustness across a large number of source distributions in CIFAR-100, and under a more specific, yet high-resolution, distribution in Imagewoof.

For our evaluation, we employed the MobileNetV2 \cite{sandler2018mobilenetv2} architecture to accommodate edge device constraints, partitioning it into an encoder $f_e(\cdot)$, comprising the first four residual blocks, and a decoder $f_d(\cdot)$ with the remaining blocks. The encoder-decoder and codebooks were jointly trained using the Adam optimizer, with  learning rate $1\cdot 10^{-4}$ and batch sizes $32$ and $16$ for CIFAR-100 and Imagewoof, respectively.

We evaluate the performance in terms of test accuracy achieved with the following quantization methods: 
$1)$ \ac{artoveq} (as detailed in Subsection~\ref{ssec:rate_adaptive_mech});
$2)$ a single-rate \ac{vqvae}, in which a different codebook is trained for each bit budget, constituting an upper-bound on the performance achievable with a single codebook shared among all resolutions;
$3)$ mixed resolution \ac{artoveq} (detailed in Subsection~\ref{ssec:identical_resolution}); 
$4)$ progressive \ac{artoveq} (detailed in Subsection~\ref{ssec:successive_refinement});
$5)$ residual \ac{vqvae} (RVQ-VAE) \cite{toderici2015variable};
and $6)$ Single-Rate \ac{lbg}, {in which \ac{lbg} \cite{linde1980algorithm} is applied anew to the learned encoder output for quantization for each codebook size}. 
In each experiment, the encoder's output was divided into $M$ segments,  and the quantizer is applied to each sub-vector.

\begin{figure*}[t]
  \centering
  \begin{subfigure}{\figWidth}
    \includegraphics[width=\linewidth]{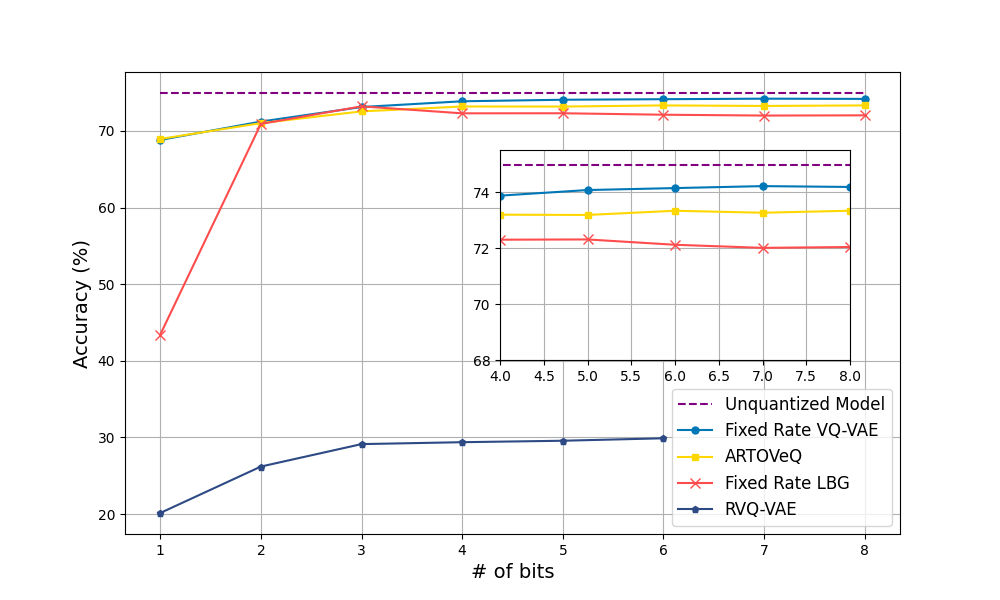}
    \caption{Accuracy vs. bits, $d=2$: ARTOVeq outperforms RVQ-VAE and LBG, closely matching fixed-rate VQ-VAE}
  \end{subfigure}
  \hfill
  \begin{subfigure}{\figWidth}
    \includegraphics[width=\linewidth]{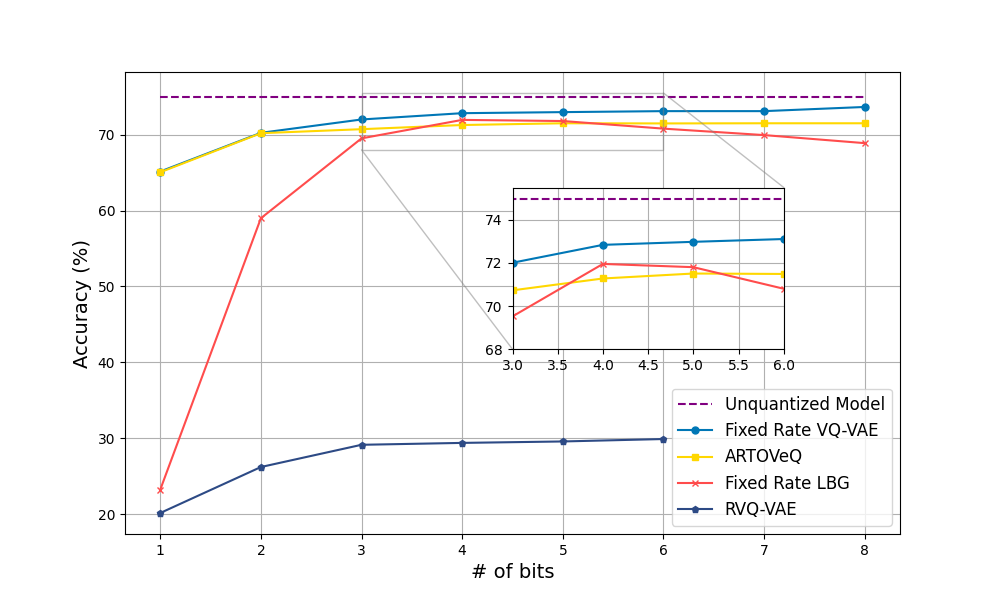}
    \caption{Accuracy vs. bits, $d=4$: ARTOVeq outperforms RVQ-VAE and LBG, while slightly underperforming fixed-rate VQ-VAE}
  \end{subfigure}

\caption{CIFAR-100 Accuracy as a function of bits per sub-vector for varying codebook vector sizes $d$.}
\label{fig:rate_adaptive_performance}
\end{figure*}

\subsection{Variable-Rate Task-Based Compression}
\label{ssec:variable_rate}

{We first assess the performance of variable-rate \ac{artoveq} in an environment where bit-rate availability may vary over time, demonstrating its capability to enable remote inference across a broad range of communication conditions.} We show that, despite utilizing a single codebook across multiple resolutions,  \ac{artoveq} remains competitive with single-rate \ac{vqvae} and outperforms other benchmark models.

\textbf{Learned Codebooks:} A defining characteristic of \ac{artoveq} is its nested codebook structure. Fig.~\ref{fig:learned_cb_vectors} illustrates the progression of codebooks $\mathcal{Q}_1,\ldots,\mathcal{Q}_8$ for $d=2$ on the CIFAR-100 dataset. As the bit resolution increases, the codebook vectors progressively capture the latent state distribution with greater precision, leading to improved performance at higher resolutions.

\textbf{Performance Evaluation:} 
Having showcased the codebook structures learned by \ac{artoveq}, we proceed to evaluating its performance when integrated into a remote inference system. The results achieved for  CIFAR-100 are reported in Fig.~\ref{fig:rate_adaptive_performance}. There, we observe the trade-offs between compression via quantization and performance in  \ac{artoveq} compared to other variable-rate and single-rate baselines. As expected, each approach exhibits an increasing trend in performance before tapering off and saturating at higher resolutions, typically around 4–5 bits for per sub-vector.

\ac{artoveq} consistently performs just slightly below the single-rate \ac{vqvae}, with a performance drop of approximately $0.8\%$  for $d=2$. Some performance degradation is observed   when transitioning from $d=2$ to $d=4$ as the number of bits per codeword is {remains constant, while the dimensionality of the codewords increases}, 
i.e., the quantizaiton rate is reduced.  Despite this,  ARTOVeQ still outperforms both single-rate \ac{lbg} and RVQ-VAE. This accuracy degradation relative to single-rate \ac{vqvae} can be attributed to the constraints imposed by ARTOVeQ’s nested codebook structure. While ARTOVeQ supports multiple bit resolutions within a single codebook, it is limited in its ability to independently optimize for each resolution, a benefit that single-rate \ac{vqvae} possesses.
The LBG algorithm consistently ranks second to last, likely due to the diverse distribution of CIFAR-100. 

\begin{figure*}
  \begin{subfigure}{\figWidth}
    \includegraphics[width=\linewidth]{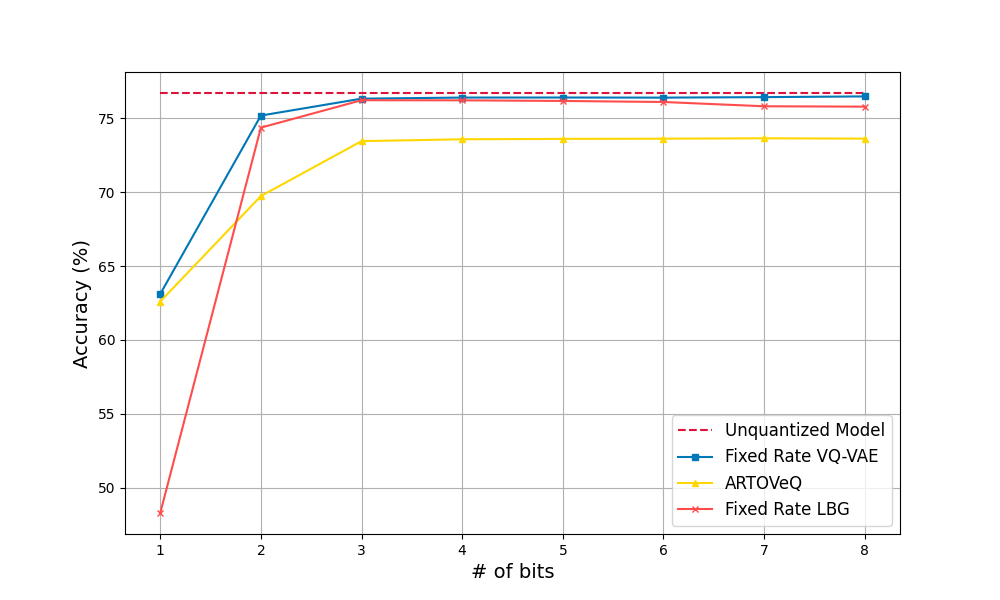}
    \caption{Accuracy vs.  bits  $d=2$: \ac{artoveq} underperforms compared to fixed-rate methods}
  \end{subfigure}
  \hfill
  \begin{subfigure}{\figWidth}
    \includegraphics[width=\linewidth]{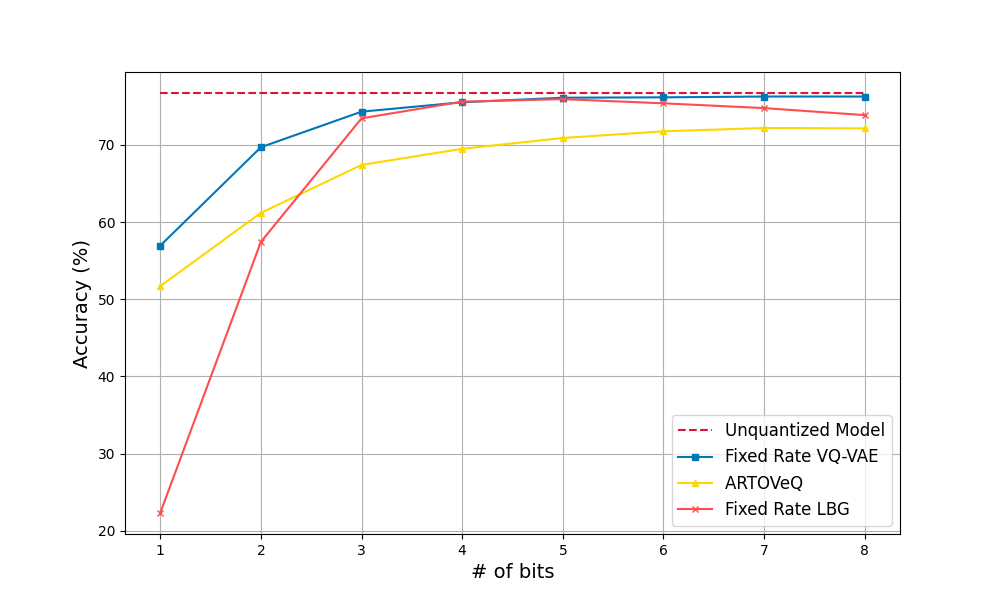}
    \caption{Accuracy vs.  bits  $d=4$: \ac{artoveq} underperforms compared to  fixed-rate methods}
  \end{subfigure}
  \caption{Imagewoof: Comparison of accuracy vs. total number of bits for different values of $d$: the unquantized MobileNetV2, the single-rate standalone \ac{vqvae}, ARTOVeQ, and LBG}
  \label{fig:imagewoof_adaptive}
\end{figure*}

Similar results are observed in the Imagewoof data, as reported in Fig.~\ref{fig:imagewoof_adaptive}. 
As seen in Fig.~\ref{fig:imagewoof_adaptive},  as opposed to CIFAR-100, here the single codebook  of \ac{artoveq} results in its reaching a small gap in performance from fixed-rate methods, that use a different codebook for each resolution. This  can be attributed to the higher redundancy at higher resolutions  of the source data, and its smaller number of labels, which allows fixed rate methods to obtain suitable codebooks for sufficient number of bits at the output of the encoder.  ARTOVeQ maintains a consistent $2\%$ performance gap, which reflects the challenge of achieving an optimal latent representation given the constraints imposed by the nested structure’s. 
RVQ-VAE, which achieves the lowest accuracy on CIFAR-100, was shown to be instable and fail to faithfully learn, and was thus not included in the figure.  
 ARTOVeQ consistently performs well across both datasets, demonstrating its robustness in diverse scenarios. 

In terms of complexity, ARTOVeQ operates in a one-shot manner, in contrast to the iterative encoding-decoding process of RVQ-VAE, which relies on residuals. As a result, ARTOVeQ offers a significant advantage in computational efficiency. In RVQ-VAE, the multiple forward pass iterations required to achieve higher resolutions substantially increase computational demand.


\subsection{Mixed-Resolution Compression}
\label{ssec:mixed_resolution}
We proceed to evaluating the ability of \ac{artoveq} to leverage its multi-resolution codebook to quantize different sub-vectors with different resolutions. 
For this task, we partitioned the encoder output, $\boldsymbol{x}_t^e$, into four blocks with a manual bit allocation strategy. The first segments were assigned the highest bit representations, following a policy where the largest bit share is allocated to the first segment, with subsequent segments receiving progressively lower bit resolutions that collectively sum to a predefined bit budget $B_t$. The aim of this study is the examine the performance of using mixed resolution codebooks compared to identical resolution ones with the same overall bit budget (which we contrasted with various benchmarks in Subsection~\ref{ssec:variable_rate}). 

\begin{figure}
    \centering
    \includegraphics[width=\figWidth]{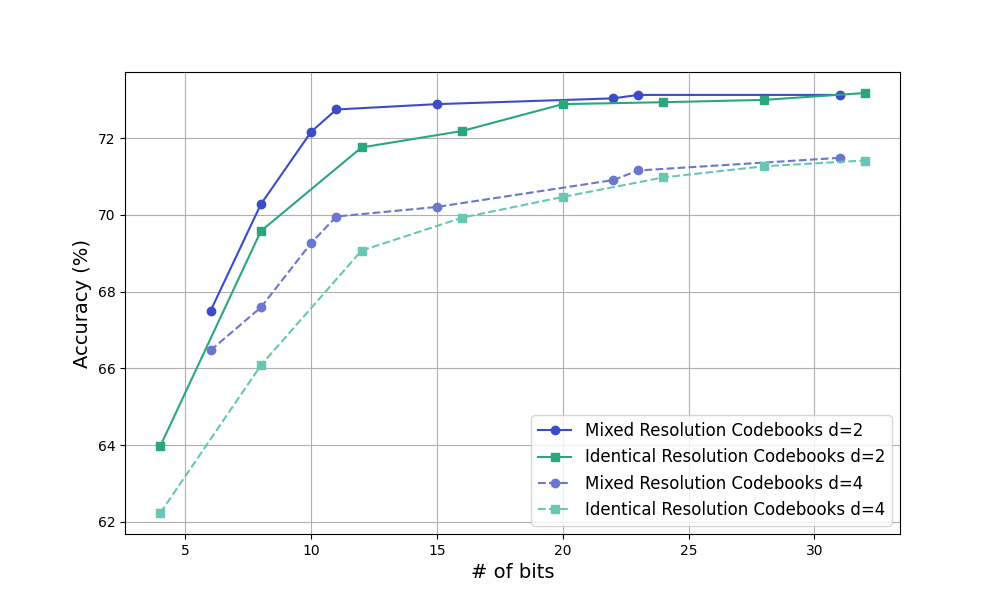}
    \caption{CIFAR-100: Accuracy versus total number of bits for four configurations—mixed resolution and identical resolution. Solid lines correspond to $d=2$, while dashed lines represent $d=4$. Mixed resolution demonstrates a broader range of allocated bits, leading to improved performance. }
    \label{fig:mixed_CIFAR}
\end{figure}

\begin{figure}
    \centering
    \includegraphics[width=\figWidth]{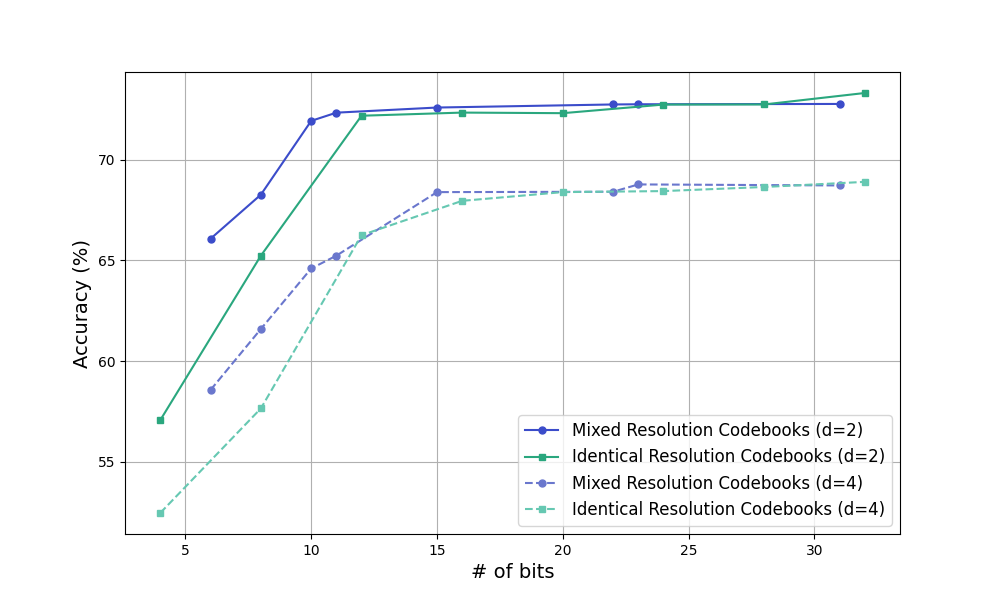}
    \caption{Imagewoof: Accuracy versus total number of bits for mixed resolution and identical resolution. Solid lines correspond to $d=2$, while dashed lines represent $d=4$. The performance of mixed resolutions closely aligns with that of identical resolutions. }
    \label{fig:imagewoof_mixed}
\end{figure}

The CIFAR-100 and Imagewoof results corresponding to $d=2$ and $d=4$ are shown in Figs.~\ref{fig:mixed_CIFAR}-\ref{fig:imagewoof_mixed}, respectively. There, we compare accuracy for different values of total number of bits assigned across four quantizers (that are applied to each four sub-vectors). 
In the identical resolution case, all quantizers have the same codebook, while in the mixed resolution case, {the first quantizer uses more codewords compared to the remaining ones. }
Our findings reveal that, for both datasets, mixed-resolution configurations consistently outperform their identical resolution counterparts, though the improvement varies with the number of bits. This demonstrates that the finer granularity enabled by mixed-resolution compression, combined with task-based learning, allows for a more refined latent space representation, resulting in improved performance. As seen in Fig.~\ref{fig:mixed_CIFAR}, this effect is particularly evident in lower bit budgets, between 4-10 bits, where the richer bit spectrum enables more effective learning and performance gains. 

Quantitatively, we observe a performance gap of approximately $0.4\% - 0.7\%$ for $d=2$ and $0.2\% - 1.5\%$ for $d=4$ on the CIFAR-100 dataset within the 8-20 bit range. Similarly, for the Imagewoof dataset, the gap ranges from $0.2\% - 3\%$ for $d=2$ and $1.6\% - 4\%$ for $d=4$. These findings suggest that improved performance can be achieved with a smaller bit budget. At the higher end of the bit spectrum, identical-resolution configurations tend to closely match the performance of mixed-resolution ones for both datasets. However, due to the redundancy and narrower distribution of Imagewoof, this alignment is reached at a lower bit budget. In all cases, mixed-resolution configurations  offer the advantage of flexible memory usage.



\subsection{Progressive Compression}
\label{ssec:progessive_comp}

We proceed by evaluating  the progressive quantization codebook version of ARTOVeQ, with its successive refinement approach. Specifically, we aim to assess how effectively the progressive constraint and its corresponding learning technique balance compression efficiency and task accuracy, and to compare the performance of successive bit increments against the variable-rate ARTOVeQ evaluated in Subsection~\ref{ssec:variable_rate}.

\begin{figure}
   \includegraphics[width=\figWidth]{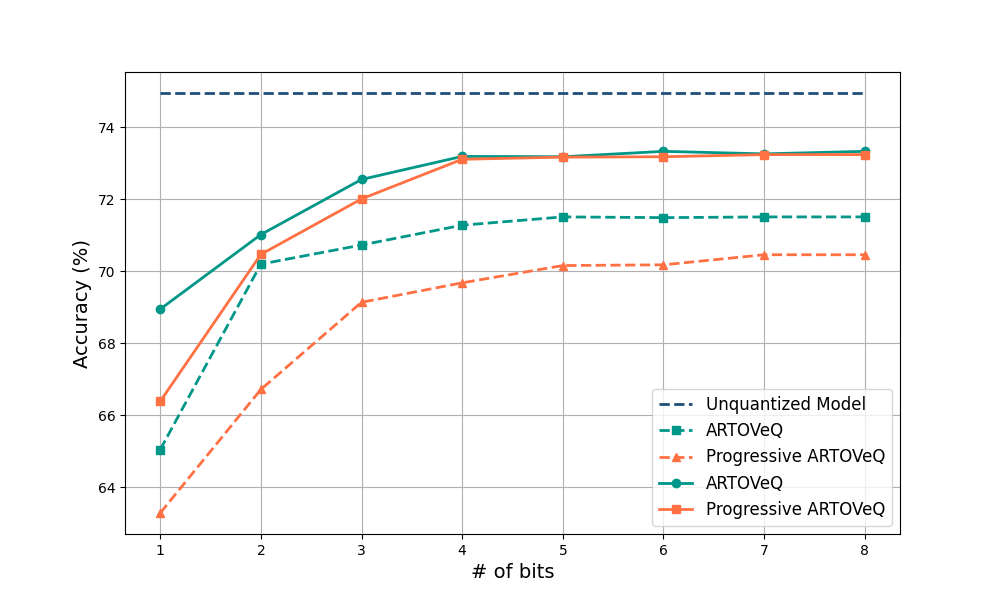}
    \caption{CIFAR-100: Comparison of accuracy as a function of the total number of bits for successive refinement and variable-rate ARTOVeQ. Solid lines indicate $d=2$, and dashed lines indicate $d=4$. Variable-rate ARTOVeQ consistently outperforms successive refinement, with the performance gap most prominent at lower bit rates (2–3 bits).  }
    \label{fig:cifar_successive_refinement}
\end{figure}

\begin{figure}
   \centering
   \includegraphics[width=\figWidth]{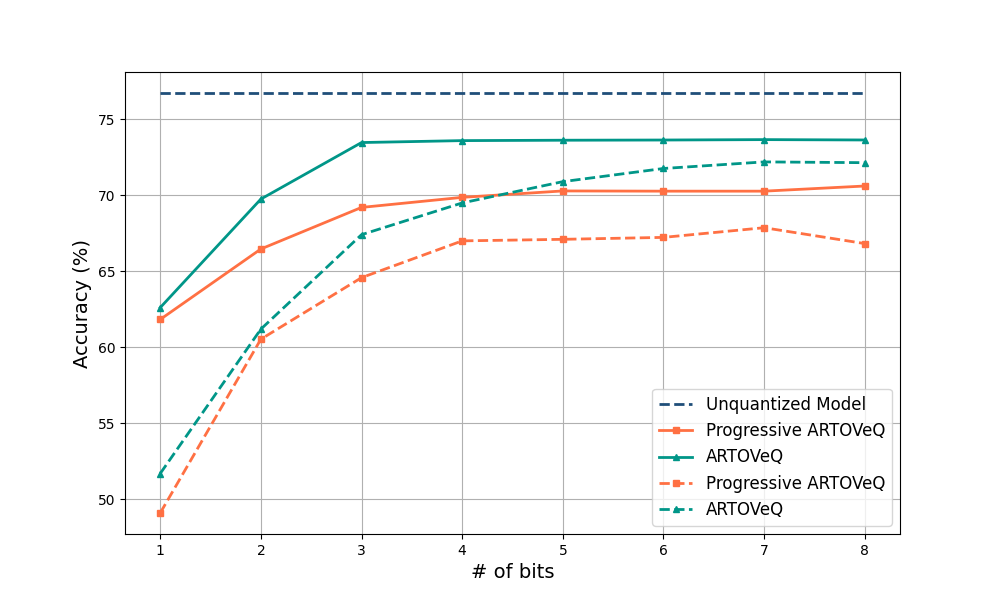}
   \caption{Imagewoof: Accuracy comparison as a function of the total number of bits for successive refinement and variable-rate ARTOVeQ. Solid lines represent $d=2$, while dashed lines represent $d=4$. ARTOVeQ demonstrates superior performance, with a consistent gap of approximately $2\%$ across all bit rates. }
    \label{fig:imagewoof_successive}
\end{figure}

Our findings, shown in Fig.~\ref{fig:cifar_successive_refinement} for CIFAR-100 and in Fig.~\ref{fig:imagewoof_successive}  for Imagewoof,  demonstrate that, as expected, variable-rate \ac{artoveq} consistently outperforms the more constrained progressive codebook  across all bit resolutions and quantization embeddings ($d=2$ and $d=4$). Nonetheless, progressive codebooks manage to  approach the performance of variable-rate ARTOVeQ owing to its dedicated learning technique, within some performance gap that varies between the considered tasks. The discrepancy is attributed to the the strict progressive constraint, which, while allowing for incremental decoding with minimal latency, comes at the cost of some performance degradation compared to variable rate \ac{artoveq}. This is particularly evident for Imagewoof with $d=4$. 


Despite the performance gap, the results show that both variable-rate ARTOVeQ and progressive \ac{artoveq} begin to saturate around 6 bits, with only marginal improvement beyond this point.   From a  complexity standpoint, both techniques operate in a one-shot fashion; however,  progressive quantization has the advantage of minimal latency, as inference can begin immediately after the first bit is received, whereas variable-rate \ac{artoveq} requires the entire bit sequence.

\subsection{Remote Inference  over Dynamic Channels}
\label{subsec:latency}
The experimental studies so far have evaluated the different versions of \ac{artoveq}, all trained to support multiple quantization resolutions, in a given bit rate. 
As the motivation for \ac{artoveq} is to facilitate remote inference over dynamic channel with a single codebook, we next evaluate its aggregated performance when repeatedly applied for remote inference with changing channel conditions. 

{\bf Experimental Setup:}
{To evaluate the performance of models in a dynamic channel environment, we consider a communication system where the channel capacities $C_t$ fluctuate over time with coherence duration $\tau$. At a given time $t$, the channel can support a maximal bit-rate $B_t = \tau \cdot C_t$, such that the per-codebook bit-budget $B_t/M$ takes values in $\{1,2\ldots,8\}$. 
To reflect variability, we simulate three distinct channel scenarios}: $S1$ a uniform distribution of bit-rates; $S2$ scenarios where lower bit-rates are more likely, and $S3$ scenarios where higher bit-rates are more likely. Theses  are obtained by setting 
\begin{equation*}
    \Pr\big(B_t = M\cdot b) = \dfrac{e^{k b}}{\sum_{b'=1}^{8}e^{k b'}}, \quad b\in \{1,2,\ldots 8\},
\end{equation*}
 where we set $k=0, -0.25, 0.25$ to obtain scenarios $S1$,  $S2$ and $S3$, respectively.




\textbf{Model Evaluation:} 
We compare the average accuracy of \ac{artoveq} and Progressive \ac{artoveq}, which both maintain a single \ac{dnn} and a single codebook, to two main benchmarks: 
The first is remote inference system that maintains eight different fixed-rate encoder-\ac{vqvae}-decoder chains, constituting the most flexible yet extremely costly alternative. We also compare to using a single fixed-rate encoder-\ac{vqvae}-decoder designed with codebook sizes $\log_2 S \in \{1,4,8\}$. As the latter operates at a fixed rate, it fails to convey the samples within the coherence time when its rate surpasses that supported by the channel.

\begin{table*}[!htbp]
\centering
\small
\renewcommand{\arraystretch}{1.3}
\caption{Performance comparison of various models for time-varying channels.}
\resizebox{0.7\linewidth}{!}
{%
\begin{tabular}{| c  l l  c c c  c c c |}
\toprule
\multirow{2}{*}{\textbf{$d$}}& \multicolumn{2}{ c }{\multirow{2}{*}{\textbf{Model}}}  &\multicolumn{3}{c}{\textbf{CIFAR-100}} &\multicolumn{3}{c|}{\textbf{ImageWoof}}  \\
 &  & & \textbf{$S1$} & \textbf{$S2$} & \textbf{$S3$} & \textbf{$S1$} & \textbf{$S2$} & \textbf{$S3$} \\
\midrule
\rowcolors{2}{white}{lightgray}
\multirow{8}{*}{\textbf{2}} & \multicolumn{2}{ c }{\textbf{Multiple Fixed-Rate \ac{vqvae}} } & 72.95 & 72.00 & 73.65 & 74.59 & 72.75 & 75.76 \\
\cline{2-9} 
& \multirow{3}{*}{\textbf{Single-Rate \ac{vqvae}}} & $1$-bit & 68.82 & 68.82 & 68.82 & 63.11 & 63.11 & 63.11 \\
\cline{3-9} 
& & $4$-bit & 46.17 & 28.79 & 60.95 & 47.75 & 29.78 & 63.05 \\
\cline{3-9} 
& & $8$-bit & 9.27 & 3.30 & 18.98 & 9.56 & 3.40 & 19.57 \\
\cline{2-9} 
& \multicolumn{2}{ c }{\textbf{\ac{artoveq}}} & 72.35 & 71.59 & 72.90 & 71.73 & 70.00 & 72.90 \\
\cline{2-9}
& \multicolumn{2}{ c }{\textbf{Progressive \ac{artoveq}}} & 71.85 & 70.72 & 72.66 & 68.59 & 67.14 & 69.64 \\
\midrule
\multirow{8}{*}{\textbf{4}} & \multicolumn{2}{ c }{\textbf{Multiple Fixed-Rate \ac{vqvae}} } & 71.64 & 70.31 & 72.61 & 72.65 & 69.59 & 74.77 \\
\cline{2-9}
& \multirow{3}{*}{\textbf{Single-Rate \ac{vqvae}}} & $1$-bit  & 65.15 & 65.15 & 65.15 & 56.94 & 56.94 & 56.94 \\
\cline{3-9} 
& & $4$-bit  & 45.53 & 28.40 & 60.11  & 47.19 & 29.43 & 62.31 \\
\cline{3-9}
& & $8$-bit  & 9.21 & 3.27 & 18.84  & 9.53 & 3.39 & 19.51 \\
\cline{2-9}
& \multicolumn{2}{ c }{\textbf{\ac{artoveq}}} & 70.41 & 69.45 & 71.07 & 67.08 & 63.52 & 69.80 \\
\cline{2-9}
& \multicolumn{2}{ c }{\textbf{Progressive \ac{artoveq}}} & 68.76 & 67.53 & 69.68 & 63.77 & 60.82 & 65.82 \\
\bottomrule
\end{tabular}%
}
\label{tab:cifar100_results}
\end{table*}

 

The results obtained with  the CIFAR-100 and the Imagewoof  dataset sare reported in Table \ref{tab:cifar100_results}.
The experimental results across CIFAR-100 and Imagewoof datasets reveal consistent performance behaviors for both $d=2$ and $d=4$. As expected, Multiple Fixed-Rate \ac{vqvae} consistently achieves the highest inference accuracy, while being only within a minor gap from  \ac{artoveq} across all scenarios. The progressive \ac{artoveq}, designed with the mechanism of incremental codebook vector's improvement, shows a slight performance drop of approximately $0.5\%-2\%$ on CIFAR-100 and $3\%-4\%$ on Imagewoof. Conversely, the Single-Rate \ac{vqvae} struggles in scenarios where the channel's supported bit-rate is insufficient to meet the model's pre-defined bit-rate requirement. This limitation highlights the lack of versatility, as it can not perform inference under constrained channel conditions. 
These results indicate on the ability of \ac{artoveq} and its  variants to support flexible remote inference over dynamic channels.

\section{Conclusions}
\label{sec:conclusions}

We proposed \ac{artoveq}, a \ac{dnn}-based remote inference mechanism with a single multi-resolution codebook that supports multi-rate vector quantization with both identical, mixed-level, and progressive resolutions. We devised a method to learn nested codebooks via a dedicated gradual learning scheme, enabling a single model to operate at various resolutions. Our numerical analyses highlight the performance trade-offs between our rate-adaptive mechanisms and model-based as well as data-driven alternatives for task-based vector quantization, showing the ability of \ac{artoveq} to learn a remote inference system with single codebook whose performance approaches systems where each rate is trained individually.

\bibliographystyle{IEEEtran}
\bibliography{IEEEabrv,refs}

\begin{thebibliography}{10}
\providecommand{\url}[1]{#1}
\csname url@samestyle\endcsname
\providecommand{\newblock}{\relax}
\providecommand{\bibinfo}[2]{#2}
\providecommand{\BIBentrySTDinterwordspacing}{\spaceskip=0pt\relax}
\providecommand{\BIBentryALTinterwordstretchfactor}{4}
\providecommand{\BIBentryALTinterwordspacing}{\spaceskip=\fontdimen2\font plus
\BIBentryALTinterwordstretchfactor\fontdimen3\font minus \fontdimen4\font\relax}
\providecommand{\BIBforeignlanguage}[2]{{%
\expandafter\ifx\csname l@#1\endcsname\relax
\typeout{** WARNING: IEEEtran.bst: No hyphenation pattern has been}%
\typeout{** loaded for the language `#1'. Using the pattern for}%
\typeout{** the default language instead.}%
\else
\language=\csname l@#1\endcsname
\fi
#2}}
\providecommand{\BIBdecl}{\relax}
\BIBdecl

\bibitem{malka2023learning}
M.~Malka, S.~Ginzach, and N.~Shlezinger, ``Learning multi-rate vector quantization for remote deep inference,'' in \emph{IEEE International Conference on Acoustics, Speech, and Signal Processing Workshops (ICASSPW)}, 2023.

\bibitem{shlezinger2022collaborative}
N.~Shlezinger and I.~V. Bajic, ``Collaborative inference for {AI}-empowered {IoT} devices,'' \emph{IEEE Internet of Things Magazine}, vol.~5, no.~4, pp. 92--98, 2022.

\bibitem{zou2022goal}
H.~Zou, C.~Zhang, S.~Lasaulce, L.~Saludjian, and H.~V. Poor, ``Goal-oriented quantization: Analysis, design, and application to resource allocation,'' \emph{{IEEE} J. Sel. Areas Commun.}, vol.~41, no.~1, pp. 42--54, 2022.

\bibitem{cover1999elements}
T.~M. Cover and J.~A. Thomas, \emph{Elements of information theory}.\hskip 1em plus 0.5em minus 0.4em\relax John Wiley \& Sons, 1999.

\bibitem{chen2024information}
J.~Chen, Y.~Fang, A.~Khisti, A.~Ozgur, N.~Shlezinger, and C.~Tian, ``Information compression in the {AI} era: Recent advances and future challenges,'' \emph{{IEEE} J. Sel. Areas Commun.}, 2025, early access.

\bibitem{shisher2024timely}
M.~K.~C. Shisher, Y.~Sun, and I.-H. Hou, ``Timely communications for remote inference,'' \emph{{IEEE/ACM} Trans. Netw.}, vol.~32, no.~5, pp. 3824--3839, 2024.

\bibitem{lu2023semantics}
Z.~Lu, R.~Li, K.~Lu, X.~Chen, E.~Hossain, Z.~Zhao, and H.~Zhang, ``Semantics-empowered communications: A tutorial-cum-survey,'' \emph{{IEEE} Commun. Surveys Tuts.}, vol.~26, no.~1, pp. 41--79, 2024.

\bibitem{shlezinger2018hardware}
N.~Shlezinger, Y.~C. Eldar, and M.~R. Rodrigues, ``Hardware-limited task-based quantization,'' \emph{{IEEE} Trans. Signal Process.}, vol.~67, no.~20, pp. 5223--5238, 2019.

\bibitem{neuhaus2021task}
P.~Neuhaus, N.~Shlezinger, M.~D{\"o}rpinghaus, Y.~C. Eldar, and G.~Fettweis, ``Task-based analog-to-digital converters,'' \emph{{IEEE} Trans. Signal Process.}, vol.~69, pp. 5403--5418, 2021.

\bibitem{agheli2022semantics}
P.~Agheli, N.~Pappas, and M.~Kountouris, ``Semantics-aware source coding in status update systems,'' in \emph{IEEE International Conference on Communications Workshops (ICC Workshops)}, 2022, pp. 169--174.

\bibitem{kountouris2021semantics}
M.~Kountouris and N.~Pappas, ``Semantics-empowered communication for networked intelligent systems,'' \emph{{IEEE} Commun. Mag.}, vol.~59, no.~6, pp. 96--102, 2021.

\bibitem{zhang2022goal}
C.~Zhang, H.~Zou, S.~Lasaulce, W.~Saad, M.~Kountouris, and M.~Bennis, ``Goal-oriented communications for the {IoT} and application to data compression,'' \emph{IEEE Internet of Things Magazine}, vol.~5, no.~4, pp. 58--63, 2022.

\bibitem{di2023goal}
P.~Di~Lorenzo, M.~Merluzzi, F.~Binucci, C.~Battiloro, P.~Banelli, E.~C. Strinati, and S.~Barbarossa, ``Goal-oriented communications for the {IoT}: System design and adaptive resource optimization,'' \emph{IEEE Internet of Things Magazine}, vol.~6, no.~4, pp. 26--32, 2023.

\bibitem{gunduz2022beyond}
D.~G{\"u}nd{\"u}z, Z.~Qin, I.~E. Aguerri, H.~S. Dhillon, Z.~Yang, A.~Yener, K.~K. Wong, and C.-B. Chae, ``Beyond transmitting bits: Context, semantics, and task-oriented communications,'' \emph{{IEEE} J. Sel. Areas Commun.}, vol.~41, no.~1, pp. 5--41, 2022.

\bibitem{Salamtian19task}
S.~Salamtian, N.~Shlezinger, Y.~C. Eldar, and M.~M{\'e}dard, ``Task-based quantization for recovering quadratic functions using principal inertia components,'' in \emph{Proc. IEEE Int. Symp. Inf. Theory}, 2019.

\bibitem{bernardo2023design}
N.~I. Bernardo, J.~Zhu, Y.~C. Eldar, and J.~Evans, ``Design and analysis of hardware-limited non-uniform task-based quantizers,'' \emph{{IEEE} Trans. Signal Process.}, vol.~71, pp. 1551--1562, 2023.

\bibitem{xie2021deep}
H.~Xie, Z.~Qin, G.~Y. Li, and B.-H. Juang, ``Deep learning enabled semantic communication systems,'' \emph{{IEEE} Trans. Signal Process.}, vol.~69, pp. 2663--2675, 2021.

\bibitem{shao2021learning}
J.~Shao, Y.~Mao, and J.~Zhang, ``Learning task-oriented communication for edge inference: An information bottleneck approach,'' \emph{{IEEE} J. Sel. Areas Commun.}, vol.~40, no.~1, pp. 197--211, 2021.

\bibitem{jankowski2020wireless}
M.~Jankowski, D.~G{\"u}nd{\"u}z, and K.~Mikolajczyk, ``Wireless image retrieval at the edge,'' \emph{{IEEE} J. Sel. Areas Commun.}, vol.~39, no.~1, pp. 89--100, 2020.

\bibitem{torfason2018towards}
R.~Torfason, F.~Mentzer, E.~Agustsson, M.~Tschannen, R.~Timofte, and L.~Van~Gool, ``Towards image understanding from deep compression without decoding,'' \emph{arXiv preprint arXiv:1803.06131}, 2018.

\bibitem{balle2020nonlinear}
J.~Ball{\'e}, P.~A. Chou, D.~Minnen, S.~Singh, N.~Johnston, E.~Agustsson, S.~J. Hwang, and G.~Toderici, ``Nonlinear transform coding,'' \emph{{IEEE} J. Sel. Topics Signal Process.}, vol.~15, no.~2, pp. 339--353, 2020.

\bibitem{shlezinger2019deep}
N.~Shlezinger and Y.~C. Eldar, ``Deep task-based quantization,'' \emph{Entropy}, vol.~23, no.~1, p. 104, 2021.

\bibitem{shlezinger2022deep}
N.~Shlezinger, A.~Amar, B.~Luijten, R.~J. van Sloun, and Y.~C. Eldar, ``Deep task-based analog-to-digital conversion,'' \emph{{IEEE} Trans. Signal Process.}, vol.~70, pp. 6021--6034, 2022.

\bibitem{danial2024power}
T.~Vol, L.~Danial, and N.~Shlezinger, ``Power-aware task-based learning of neuromorphic {ADC}s,'' in \emph{IEEE International Conference on Acoustics, Speech and Signal Processing (ICASSP)}, 2024, pp. 5850--5854.

\bibitem{van2017neural}
A.~Van Den~Oord and O.~Vinyals, ``Neural discrete representation learning,'' \emph{Advances in Neural Information Processing Systems}, vol.~30, 2017.

\bibitem{malka2022decentralized}
M.~Malka, E.~Farhan, H.~Morgenstern, and N.~Shlezinger, ``Decentralized low-latency collaborative inference via ensembles on the edge,'' \emph{{IEEE} Trans. Wireless Commun.}, 2024, early access.

\bibitem{mishra2022deep}
D.~Mishra, S.~K. Singh, and R.~K. Singh, ``Deep architectures for image compression: a critical review,'' \emph{Signal Processing}, vol. 191, p. 108346, 2022.

\bibitem{habibian2019video}
A.~Habibian, T.~v. Rozendaal, J.~M. Tomczak, and T.~S. Cohen, ``Video compression with rate-distortion autoencoders,'' in \emph{Proceedings of the IEEE/CVF International Conference on Computer Vision}, 2019, pp. 7033--7042.

\bibitem{zeghidour2021soundstream}
N.~Zeghidour, A.~Luebs, A.~Omran, J.~Skoglund, and M.~Tagliasacchi, ``Soundstream: An end-to-end neural audio codec,'' \emph{{IEEE} Trans. Acoust., Speech, Signal Process.}, vol.~30, pp. 495--507, 2021.

\bibitem{yang2023introduction}
Y.~Yang, S.~Mandt, L.~Theis \emph{et~al.}, ``An introduction to neural data compression,'' \emph{Foundations and Trends{\textregistered} in Computer Graphics and Vision}, vol.~15, no.~2, pp. 113--200, 2023.

\bibitem{li2018auto}
G.~Li, L.~Liu, X.~Wang, X.~Dong, P.~Zhao, and X.~Feng, ``Auto-tuning neural network quantization framework for collaborative inference between the cloud and edge,'' in \emph{International Conference on Artificial Neural Networks (ICANN)}.\hskip 1em plus 0.5em minus 0.4em\relax Springer, 2018, pp. 402--411.

\bibitem{agustsson2017soft}
E.~Agustsson, F.~Mentzer, M.~Tschannen, L.~Cavigelli, R.~Timofte, L.~Benini, and L.~V. Gool, ``Soft-to-hard vector quantization for end-to-end learning compressible representations,'' in \emph{Advances in Neural Information Processing Systems}, 2017, pp. 1141--1151.

\bibitem{cai2019efficient}
C.~Cai, L.~Chen, X.~Zhang, and Z.~Gao, ``Efficient variable rate image compression with multi-scale decomposition network,'' \emph{{IEEE} Trans. Circuits Syst. Video Technol.}, vol.~29, no.~12, pp. 3687--3700, 2019.

\bibitem{choi2019variable}
Y.~Choi, M.~El-Khamy, and J.~Lee, ``Variable rate deep image compression with a conditional autoencoder,'' in \emph{IEEE/CVF International Conference on Computer Vision (ICCV)}, 2019, pp. 3146--3154.

\bibitem{yang2020variable}
F.~Yang, L.~Herranz, J.~Van De~Weijer, J.~A.~I. Guiti{\'a}n, A.~M. L{\'o}pez, and M.~G. Mozerov, ``Variable rate deep image compression with modulated autoencoder,'' \emph{{IEEE} Signal Process. Lett.}, vol.~27, pp. 331--335, 2020.

\bibitem{yang2021slimmable}
F.~Yang, L.~Herranz, Y.~Cheng, and M.~G. Mozerov, ``Slimmable compressive autoencoders for practical neural image compression,'' in \emph{IEEE/CVF Conference on Computer Vision and Pattern Recognition}, 2021, pp. 4998--5007.

\bibitem{lee2022selective}
J.~Lee, S.~Jeong, and M.~Kim, ``Selective compression learning of latent representations for variable-rate image compression,'' \emph{Advances in Neural Information Processing Systems}, vol.~35, pp. 13\,146--13\,157, 2022.

\bibitem{gupta2022user}
R.~Gupta, S.~BV, N.~Kapoor, R.~Jaiswal, S.~R. Nangi, and K.~Kulkarni, ``User-guided variable rate learned image compression,'' in \emph{IEEE/CVF Conference on Computer Vision and Pattern Recognition}, 2022, pp. 1753--1758.

\bibitem{duan2024qarv}
Z.~Duan, M.~Lu, J.~Ma, Y.~Huang, Z.~Ma, and F.~Zhu, ``{QARV}: Quantization-aware {R}es{N}et {VAE} for lossy image compression,'' \emph{{IEEE} Trans. Pattern Anal. Mach. Intell.}, vol.~46, no.~1, pp. 436--450, 2024.

\bibitem{seo2024raqvae}
J.~Seo and J.~Kang, ``{RAQ-VAE}: Rate-adaptive vector-quantized variational autoencoder,'' \emph{arXiv preprint arXiv:2405.14222}, 2024.

\bibitem{toderici2015variable}
G.~Toderici, S.~M. O'Malley, S.~J. Hwang, D.~Vincent, D.~Minnen, S.~Baluja, M.~Covell, and R.~Sukthankar, ``Variable rate image compression with recurrent neural networks,'' \emph{arXiv preprint arXiv:1511.06085}, 2015.

\bibitem{toderici2017full}
G.~Toderici, D.~Vincent, N.~Johnston, S.~Jin~Hwang, D.~Minnen, J.~Shor, and M.~Covell, ``Full resolution image compression with recurrent neural networks,'' in \emph{IEEE Conference on Computer Vision and Pattern Recognition}, 2017, pp. 5306--5314.

\bibitem{johnston2018improved}
N.~Johnston, D.~Vincent, D.~Minnen, M.~Covell, S.~Singh, T.~Chinen, S.~J. Hwang, J.~Shor, and G.~Toderici, ``Improved lossy image compression with priming and spatially adaptive bit rates for recurrent networks,'' in \emph{IEEE Conference on Computer Vision and Pattern Recognition}, 2018, pp. 4385--4393.

\bibitem{lee2022dpict}
J.-H. Lee, S.~Jeon, K.~P. Choi, Y.~Park, and C.-S. Kim, ``{DPICT}: Deep progressive image compression using trit-planes,'' in \emph{IEEE/CVF conference on Computer Vision and Pattern Recognition}, 2022, pp. 16\,113--16\,122.

\bibitem{hojjat2023progdtd}
A.~Hojjat, J.~Haberer, and O.~Landsiedel, ``Prog{DTD}: Progressive learned image compression with double-tail-drop training,'' in \emph{Proceedings of the IEEE/CVF Conference on Computer Vision and Pattern Recognition}, 2023, pp. 1130--1139.

\bibitem{abdi2019nested}
A.~Abdi and F.~Fekri, ``Nested dithered quantization for communication reduction in distributed training,'' \emph{arXiv preprint arXiv:1904.01197}, 2019.

\bibitem{equitz1991successive}
W.~H. Equitz and T.~M. Cover, ``Successive refinement of information,'' \emph{{IEEE} Trans. Inf. Theory}, vol.~37, no.~2, pp. 269--275, 1991.

\bibitem{cheng2005successive}
S.~Cheng and Z.~Xiong, ``Successive refinement for the {W}yner-{Z}iv problem and layered code design,'' \emph{{IEEE} Trans. Signal Process.}, vol.~53, no.~8, pp. 3269--3281, 2005.

\bibitem{du2021progressive}
R.~Du, J.~Xie, Z.~Ma, D.~Chang, Y.-Z. Song, and J.~Guo, ``Progressive learning of category-consistent multi-granularity features for fine-grained visual classification,'' \emph{{IEEE} Trans. Pattern Anal. Mach. Intell.}, vol.~44, no.~12, pp. 9521--9535, 2021.

\bibitem{sandler2018mobilenetv2}
M.~Sandler, A.~Howard, M.~Zhu, A.~Zhmoginov, and L.-C. Chen, ``Mobilenetv2: Inverted residuals and linear bottlenecks,'' in \emph{IEEE Conference on Computer Vision and Pattern Recognition}, 2018, pp. 4510--4520.

\bibitem{gray1998quantization}
R.~M. Gray and D.~L. Neuhoff, ``Quantization,'' \emph{{IEEE} Trans. Inf. Theory}, vol.~44, no.~6, pp. 2325--2383, 1998.

\bibitem{kang2017neurosurgeon}
Y.~Kang, J.~Hauswald, C.~Gao, A.~Rovinski, T.~Mudge, J.~Mars, and L.~Tang, ``Neurosurgeon: Collaborative intelligence between the cloud and mobile edge,'' \emph{ACM SIGARCH Computer Architecture News}, vol.~45, no.~1, pp. 615--629, 2017.

\bibitem{gautam2023soft}
T.~Gautam, R.~Pryzant, Z.~Yang, C.~Zhu, and S.~Sojoudi, ``Soft convex quantization: Revisiting vector quantization with convex optimization,'' \emph{arXiv preprint arXiv:2310.03004}, 2023.

\bibitem{fifty2024restructuring}
C.~Fifty, R.~G. Junkins, D.~Duan, A.~Iger, J.~W. Liu, E.~Amid, S.~Thrun, and C.~R{\'e}, ``Restructuring vector quantization with the rotation trick,'' \emph{arXiv preprint arXiv:2410.06424}, 2024.

\bibitem{huang2023towards}
M.~Huang, Z.~Mao, Z.~Chen, and Y.~Zhang, ``Towards accurate image coding: Improved autoregressive image generation with dynamic vector quantization,'' in \emph{Proceedings of the IEEE/CVF Conference on Computer Vision and Pattern Recognition}, 2023, pp. 22\,596--22\,605.

\bibitem{dong2023peco}
X.~Dong, J.~Bao, T.~Zhang, D.~Chen, W.~Zhang, L.~Yuan, D.~Chen, F.~Wen, N.~Yu, and B.~Guo, ``Pe{C}o: Perceptual codebook for {BERT} pre-training of vision transformers,'' in \emph{Proceedings of the AAAI Conference on Artificial Intelligence}, vol.~37, no.~1, 2023, pp. 552--560.

\bibitem{linde1980algorithm}
Y.~Linde, A.~Buzo, and R.~Gray, ``An algorithm for vector quantizer design,'' \emph{{IEEE} Trans. Commun.}, vol.~28, no.~1, pp. 84--95, 1980.

\bibitem{shusterman1994image}
E.~Shusterman and M.~Feder, ``Image compression via improved quadtree decomposition algorithms,'' \emph{{IEEE} Trans. Image Process.}, vol.~3, no.~2, pp. 207--215, 1994.

\bibitem{zamir2002nested}
R.~Zamir, S.~Shamai, and U.~Erez, ``Nested linear/lattice codes for structured multiterminal binning,'' \emph{{IEEE} Trans. Inf. Theory}, vol.~48, no.~6, pp. 1250--1276, 2002.

\bibitem{alvar2021scalable}
S.~R. Alvar and I.~V. Baji{\'c}, ``Scalable privacy in multi-task image compression,'' in \emph{IEEE International Conference on Visual Communications and Image Processing (VCIP)}, 2021.

\bibitem{lang2023joint}
N.~Lang, E.~Sofer, T.~Shaked, and N.~Shlezinger, ``Joint privacy enhancement and quantization in federated learning,'' \emph{{IEEE} Trans. Signal Process.}, vol.~71, pp. 295--310, 2023.

\bibitem{esfahanizadeh2023infoshape}
H.~Esfahanizadeh, W.~Wu, M.~Ghobadi, R.~Barzilay, and M.~M{\'e}dard, ``Infoshape: Task-based neural data shaping via mutual information,'' in \emph{IEEE International Conference on Acoustics, Speech and Signal Processing (ICASSP)}, 2023.

\end{thebibliography}

\end{document}